\documentclass[amsmath, amssymb, aps, prx, twocolumn]{revtex4-2}
\usepackage{subcaption}
\usepackage{cancel}
\usepackage{algorithmic}
\usepackage{booktabs}
\usepackage{makecell}
\usepackage{amsthm}

\usepackage{caption}
\usepackage{graphicx}
\usepackage{dcolumn}
\usepackage{bm}
\usepackage{array}
\usepackage{float}
\usepackage{mathrsfs}
\usepackage{multirow}
\usepackage{varwidth}
\usepackage{amsmath}
\usepackage{amssymb}
\usepackage{stmaryrd}
\usepackage{hyperref}
\usepackage[version = 4]{mhchem} 
\usepackage{algorithm}
\usepackage{algorithmic}
\usepackage{chemformula}
\usepackage{siunitx}

\begin{document}

\title{Constructing Compact ADAPT Unitary Coupled-Cluster Ansatz with Parameter-Based Criterion}
\author{Runhong He$^{1}$, Xin Hong$^{1}$, Qiaozhen Chai$^{1}$, Ji Guan$^{1}$, Junyuan Zhou$^{2}$, Arapat Ablimit$^{3}$, Guolong Cui$^{4}$}
\author{Shenggang Ying $^{1}$}
\email{yingsg@ios.ac.cn}
\address{
1. Key Laboratory of System Software (Chinese Academy of Sciences), Institute of Software, Chinese Academy of Sciences, Beijing 100190, China.\\
2. MindSpore Quantum Special Interest Group, China.\\
3. College of Physical Science and Technology, Xinjiang University, Urumqi 830017, China.\\
4. Arclight Quantum Computing Inc., Beijing 100086, China.\\
}

\begin{abstract}

The adaptive derivative-assembled pseudo-trotter variational quantum eigensolver (ADAPT-VQE) is a promising hybrid quantum-classical algorithm for molecular ground state energy calculation, yet its practical scalability is hampered by redundant excitation operators and excessive measurement costs. To address these challenges, we propose Param-ADAPT-VQE, a novel improved algorithm that selects excitation operators based on a parameter-based criterion instead of the traditional gradient-based metric. This strategy effectively eludes redundant operators. We further develop a sub-Hamiltonian technique and integrate a hot-start VQE optimization strategy, achieving a significant reduction in measurement costs. Numerical experiments on typical molecular systems demonstrate that Param-ADAPT-VQE outperforms the original ADAPT-VQE in computational accuracy, ansatz size, and measurement costs. Furthermore, our scheme retains the fundamental framework of ADAPT-VQE and is thus fully compatible with its various modified versions, enabling further performance improvements in specific aspects. This work presents an efficient and scalable enhancement to ADAPT-VQE, mitigating the core obstacles that impede its practical implementation in the field of molecular quantum chemistry.
\end{abstract}

\maketitle

\section{Introduction}\label{sec:Introduction}

The accurate calculation of molecular energy levels lies at the heart
of quantum chemistry \cite{szabo}, as it yields critical insights
into chemical reactivity, material properties, and molecular dynamics
\cite{molecular_property_calculations}. However, classical computational
approaches, such as the full configuration interaction (FCI), exhibit
inherent limitations in addressing large molecular systems and strongly
correlated systems, arising from the exponential scaling
of computational complexity with increasing system size \cite{Quantum_computational_chemistry}.

The emergence of quantum computing \cite{qc_nielsen} offers a promising
solution to this challenge,  and the variational quantum eigensolver (VQE) algorithm \cite{VQE_first,vqe_review,vqe_review2} is among the most promising and widely investigated approaches for future quantum chemistry applications.
VQE leverages the variational principle \cite{RRvariational}
to find the ground state energy of a target system by minimizing
the expectation value of its Hamiltonian with respect to a parameterized
quantum state (ansatz). The key advantage of VQE lies in that it splits
the computational task between quantum and classical devices: the
quantum processor prepares the ansatz and measures the Hamiltonian
expectation value, while the classical optimizer adjusts the variational
parameters iteratively. This hybrid quantum-classical architecture
renders VQE compatible with near-term noisy intermediate-scale quantum
(NISQ) devices \cite{NISQ}, which suffer from non-negligible gate
errors, environmental noise, and a limited number of available qubits.

Despite its practicality, the performance of VQE is highly dependent
on the design of the ansatz. A well-constructed ansatz should be expressive
enough to approximate the ground state of the target system accurately
while at the same time remaining shallow enough to be implementable
on NISQ devices \cite{Quantum_computational_chemistry}.
Traditional fixed ansatze (e.g., unitary coupled cluster singles and doubles, UCCSD \cite{ucc_review,VQE_point_group_symmetry,HiUCCSD}) often require a large number of excitation operators to achieve sufficient accuracy for complex molecular systems, thereby increasing noise susceptibility and exacerbating training challenges.

To address this limitation, the adaptive derivative-assembled pseudo-trotter
(ADAPT) VQE \cite{ADAPT-VQE} was developed as a problem-driven ansatz
construction method. 
It builds the ansatz incrementally by
selecting only the excitation operator with maximum
initial gradient from a pre-defined operator pool in each iteration. 
Furthermore, ADAPT-VQE has been shown to mitigate optimization challenges
associated with barren plateaus \cite{Barren_plateaus} and local
minima \cite{ADAPT_VQE_vs_bp}. ADAPT-VQE and its modified versions
\cite{ADAPT-VQE-TETRIS,ADAPT_amp_reording_batch,adapt_pool_size,adapt_symmetry_breaking,ADAPT_VQE_CEO,ADAPT_VQE_ES,ADAPT_VQE_pauli_linear_pool,ADAPT_VQE_Pruned,ADAPT_VQE_QEB,ADAPT_VQE_qubit}
have been widely applied to diverse molecular systems and have demonstrated
superior performance in balancing ansatz expressivity and depth.

Nevertheless, ADAPT-VQE still faces two key challenges: (1) The gradient-based criterion is not infallible. Several studies \cite{ADAPT_VQE_ES,ADAPT_VQE_QEB, ADAPT_VQE_Pruned} have noted that many excitation operators with the largest initial gradient are subsequently identified as redundant ones that fail to induce an effective energy reduction after optimization. They \cite{ADAPT_VQE_ES,ADAPT_VQE_QEB} resort to an energy-based criterion, but incur a quartic additional measurement cost.
(2) The compactness of the ADAPT-VQE ansatz comes at the cost of a surge in measurement overhead.
While measurement cost is not
the critical bottleneck at the current stage, excessively high measurement
cost will still restrict the application to larger molecules \cite{ADAPT_VQE_reduced_RDM,ADAPT-VQE}.

To address these challenges, we propose the Param-ADAPT-VQE algorithm
with three core improvements: a novel parameter-based excitation operator
selection criterion to reduce ansatz redundancy, a sub-Hamiltonian
technique to suppress the measurement cost surge from operator pool
scanning, and a hot-start training strategy to cut the measurement
cost of global VQE optimization. Benchmark experiments on a variety
of molecular systems demonstrate that, compared with ADAPT-VQE, Param-ADAPT-VQE
exhibits superior performance in terms of the number of excitation
operators, computational accuracy, and measurement cost.

The remainder of this paper is organized as follows: In Subsection
\ref{subsec:VQE-and-ADAPT-VQE}, we introduce the fundamentals of
the VQE and ADAPT-VQE algorithms. In Subsection \ref{subsec:Param-ADAPT-VQE-Algorithm},
we elaborate on the proposed Param-ADAPT-VQE algorithm in detail.
Numerical experiments are performed in Section \ref{sec:Numberical-Results},
and our conclusions are presented in Section \ref{sec:Conclusion}.
Additional performance comparisons for 6 typical molecules at varying bond lengths are provided in Appendix \ref{appendix} for further reference.

\section{Methods}\label{sec:Methods}

\subsection{VQE and ADAPT-VQE }\label{subsec:VQE-and-ADAPT-VQE}

In the present work, indices $i,j,k,\dots$ label occupied spin orbitals, while $a,b,c,\dots$ denote virtual spin orbitals. Spin orbitals of both types are indexed by $p,q,r,s$. The total number of spin orbitals involved in a system
of interest is denoted as $N$.

Under the Born-Oppenheimer approximation \cite{szabo},
where the molecular nuclei are treated as stationary, the second-quantized
electronic Hamiltonian of a molecule in atomic units is given by the
following form \cite{Quantum_computational_chemistry}:
\begin{equation}
\hat{H}=\sum_{p,q}^{N}h_{q}^{p}\hat{a}_{p}^{\dagger}\hat{a}_{q}+\frac{1}{2}\sum_{p,q,r,s}^{N}h_{rs}^{pq}\hat{a}_{p}^{\dagger}\hat{a}_{q}^{\dagger}\hat{a}_{r}\hat{a}_{s},\label{eq:full_ham}
\end{equation}
where $\hat{a}_{i}^{\dagger}$ and $\hat{a}_{i}$ are the fermionic
creation and annihilation operators, respectively. The electron integrals
$h_{q}^{p}$ and $h_{rs}^{pq}$ can be computed classically within a
specified basis set, such as the minimal STO-3G \cite{szabo}. The
fermionic Hamiltonian (\ref{eq:full_ham}) can be mapped to Pauli
strings via transformations such as the Jordan-Wigner \cite{Jordan-Wigner},
and its expectation value in a trial state $|\psi(\boldsymbol{\theta})\rangle$
can then be derived through measurement.

The trial state $|\psi(\boldsymbol{\theta})\rangle$ is prepared from a reference
state $|\psi_{\text{ref}}\rangle$ via a moderately deep parameterized quantum
circuit (ansatz) $U(\boldsymbol{\theta})$, such that $|\psi(\boldsymbol{\theta})\rangle=U(\boldsymbol{\theta})|\psi_{\text{ref}}\rangle$.
The reference state $|\psi_{\text{ref}}\rangle$ is usually chosen as the Hartree-Fock
state $|\text{HF}\rangle$. The measurement cost is proportional to
the number of terms in the fermionic Hamiltonian (\ref{eq:full_ham}),
scaling as $O(N^{4})$.

Within the VQE framework, an upper limit for the unknown ground-state
energy $E_{0}$ can be derived by minimizing the Hamiltonian expectation
value with respect to the variational parameters $\boldsymbol{\theta}$,
according to the Rayleigh-Ritz variational principle \cite{RRvariational}
\begin{equation}
\langle\psi(\boldsymbol{\theta})|H|\psi(\boldsymbol{\theta})\rangle\geq E_{0}.
\end{equation}

The UCCSD is a commonly used chemistry-inspired ansatz which includes
only single- and double-excitation operators to reduce the circuit
depth, i.e.,
\begin{equation}
U(\boldsymbol{\theta})=\text{e}^{\hat{T}(\boldsymbol{\theta})},
\end{equation}
where the anti-Hermitian operator
\begin{equation}
\hat{T}(\boldsymbol{\theta})=\hat{T}_{1}(\boldsymbol{\theta})+\hat{T}_{2}(\boldsymbol{\theta}),
\end{equation}
\begin{equation}
\hat{T}_{1}(\boldsymbol{\theta})=\sum_{i,a}\theta_{i}^{a}\hat{\tau}_{i}^{a}=\sum_{i,a}\theta_{i}^{a}\left(\hat{a}_{a}^{\dagger}\hat{a}_{i}-\hat{a}_{i}^{\dagger}\hat{a}_{a}\right),
\end{equation}
\begin{equation}
\hat{T}_{2}(\boldsymbol{\theta})\!=\!\!\!\!\sum_{a>b,i>j}\!\!\!\!\!\theta_{ij}^{ab}\hat{\tau}_{ij}^{ab}\!=\!\!\!\!\!\sum_{a>b,i>j}\!\!\!\!\!\theta_{ij}^{ab}\!\left(\hat{a}_{a}^{\dagger}\hat{a}_{b}^{\dagger}\hat{a}_{i}\hat{a}_{j}\!-\!\hat{a}_{i}^{\dagger}\hat{a}_{j}^{\dagger}\hat{a}_{a}\hat{a}_{b}\right).
\end{equation}
Generally, the UCCSD ansatz contains $O\left(\binom{N/2}{1}^{2}\right)=O(N^{2})$
single- and $O\!\left(\binom{N/2}{2}^{2}\right)\!\!=\!O(N^{4})$ double-excitation
operators, resulting in an overall $O(N^{4})$ complexity \cite{ucc_review,HiUCCSD}.

Within the first-order Trotter-Suzuki approximation \cite{trotter},
the above evolution operator can be decomposed as
\begin{equation}
\text{e}^{T(\boldsymbol{\theta})}\approx\prod_{i,a}\text{e}^{\hat{\tau}_{i}^{a}}\cdot\!\!\prod_{i>j,a>b}\!\!\!\text{e}^{\hat{\tau}_{ij}^{ab}}.
\end{equation}
Since variational optimization can absorb the majority of truncation
errors \cite{VQE_trotter_one}, this approximation is generally acceptable
in VQE.

The circuit implementation of excitation operators has been extensively
investigated in existing literature \cite{ADAPT_VQE_CEO,VQE_softhardware_cooptimization_ligushu,VQE_reservoir,ADAPT_VQE_qubit,adapt_9_cnot}.
Among these studies, the method proposed in \cite{efficient_excitations_circ}
has attracted widespread attention due to its remarkably shallow circuit
depth, and we adopt this efficient circuit construction approach for
all excitation operators in our work as well.

In contrast to the general-purpose and fixed ansatze, ADAPT-VQE constructs
a problem-tailored ansatz on the fly. It defines an excitation operator
pool (typically based on the UCCSD) and builds a compact quantum circuit
from the identity operator by appending only the excitation operator with the largest initial gradient magnitude from
the pool at each iteration. Given that the measurement cost for computing
the initial gradient of each operator is $O(N^{3})$, for an operator
pool with $O(N^{4})$ size, the additional quantum measurements required
to identify the proper excitation operator in each iteration scale as
$O(N^{7})$ \cite{ADAPT-VQE}. Once the selected excitation operator
is added to the ansatz, the algorithm executes a global VQE optimization
on the updated ansatz with a warm start, i.e., the converged variational
parameters from the preceding iteration act as the initial values
for the current optimization and the parameter of the new operator
is zero-initialized, enabling accelerated convergence by leveraging
the pre-optimized solution \cite{ADAPT_VQE_QEB}. This entire iterative cycle, consisting
of excitation operator pool gradient evaluation, operator
selection, and warm-started VQE optimization, is repeated until the
pool's gradient norm drops below a pre-defined convergence threshold. 

Theoretically, ADAPT-VQE can yield increasingly improved precision as iterative optimization progresses \cite{ADAPT-VQE,ADAPT_VQE_vs_bp}.
However, several studies \cite{ADAPT_VQE_Pruned,ADAPT_VQE_ES,ADAPT_VQE_QEB} have noted that ADAPT-VQE may incorporate numerous redundant operators in the ansatz, 
implying that the gradient-based selection criterion may not be optimal. Furthermore, the excessively high measurement
cost also restricts its applicability to larger-scale problems. Next,
we introduce our Param-ADAPT-VQE algorithm to mitigate these challenges.

\subsection{Param-ADAPT-VQE }\label{subsec:Param-ADAPT-VQE-Algorithm}

The general workflow of the Param-ADAPT-VQE algorithm is as follows:

Step 1. Define an excitation operator pool $\mathbb{P}\!\!:=\!\!\{\hat{\tau}_{q}^{p},\hat{\tau}_{rs}^{pq}\}$
and construct the full molecular fermionic Hamiltonian $\hat{H}$.
For each excitation operator $\hat{\tau}_{i}\in\mathbb{P}$, extract
the sub-Hamiltonian $\hat{H}_{i}$ by selecting terms sharing indices
with $\hat{\tau}_{i}$. Define iteration termination conditions: a parameter magnitude threshold $\epsilon>0$ or a maximum iteration count $k_{\text{max}}\geq0$.
Choose a suitable reference state $|\psi^{(0)}\rangle$ (e.g., the
Hartree-Fock state $|\text{HF}\rangle$), initialize the ansatz as
the identity operator $U\leftarrow U^{(0)}=\hat{I}$ and set the iteration
index to $k=1$. 

Step 2. If $k>k_{\text{max}}$, terminate the iterative procedure
and proceed to Step 6; otherwise start the $k^{\text{th}}$ iteration
as follows: prepare the state $|\psi^{(k-1)}\rangle=U(\boldsymbol{\theta}^{(k-1)*})|\psi^{(0)}\rangle$,
where $\boldsymbol{\theta}^{(k-1)*}$ denotes the optimized parameters
from the $(k-1)^{\text{th}}$ iteration. We note that in this algorithm,
the superscript $^{*}$ denotes the optimized result, rather than the
complex conjugate. For each excitation operator $\hat{\tau}_{i}\in\mathbb{P}$, 
perform a local VQE optimization to minimize the energy expectation
value of the sub-Hamiltonian $\hat{H}_{i}$ with respect to a new
parameter $\theta_{i}$ until convergence, yielding the optimal parameter
$\theta_{i}^{*}$. The local VQE optimization is defined as:
\begin{equation}
\theta_{i}^{*}=\underset{\theta_{i}}{\text{argmin}}\langle\psi^{(k-1)*}|\text{e}^{\theta_{i}\hat{\tau}_{i}^{\dagger}}\hat{H}_{i}\text{e}^{\theta_{i}\hat{\tau}_{i}}|\psi^{(k-1)*}\rangle.
\end{equation}

Step 3. If the maximum magnitude $|\theta_{i}^{*}|<\epsilon$ , terminate
the iterative procedure and proceed to Step 6.

Step 4. Select the excitation operator $\hat{\tau}_{i}$ with the
maximum magnitude $|\theta_{i}^{*}|$, and append its corresponding unitary
operator $\text{e}^{\theta_{i}^{*}\hat{\tau}_{i}}$ to the left of
the current ansatz: $U(\boldsymbol{\theta}^{(k)})\leftarrow\text{e}^{\theta_{i}^{*}\hat{\tau}_{i}}U(\boldsymbol{\theta}^{(k-1)*})$.
Update the ansatz parameters set as $\boldsymbol{\theta}^{(k)}\leftarrow\boldsymbol{\theta}^{(k-1)*}\cup\{\theta_{i}^{*}\}$. 

Step 5. Perform a\textbf{ }global VQE optimization to re-optimize
all variational parameters in the updated ansatz:
\begin{equation}
\boldsymbol{\theta}^{(k)*}=\underset{\boldsymbol{\theta}^{(k)}}{\text{argmin}}\langle\psi^{(0)}|U^{\dagger}(\boldsymbol{\theta}^{(k)})\hat{H}U(\boldsymbol{\theta}^{(k)})|\psi^{(0)}\rangle.
\end{equation}
Increment the iteration index as $k\leftarrow k+1$ and return to
Step 2.

Step 6. Evaluate the full Hamiltonian expectation value of the final
trial state $|\psi^{(k)}\rangle=U(\boldsymbol{\theta}^{(k)*})|\psi^{(0)}\rangle$,
i.e., $\langle\psi^{(k)}|\hat{H}|\psi^{(k)}\rangle$, and output this value
as the molecular ground-state energy estimate.

We now provide some more detailed comments on the steps of the above
Param-ADAPT-VQE protocol.

As described in Step 2, in a new iteration, Param-ADAPT-VQE tests
each potential excitation operator $\hat{\tau}_{i}$ in the pool individually
for the current trial state by assigning it a parameter $\theta_{i}$
and optimizing $\theta_{i}$ to convergence (i.e., local VQE optimization)
to obtain the optimized $\theta_{i}^{*}$; the operator with the largest
parameter magnitude $|\theta_{i}^{*}|$ is then selected and added to the
ansatz. \textit{We note that, although this strategy involves more additional
optimization steps than the gradient-based ADAPT-VQE, it does not
actually incur a significant increase in experimental cost --- and
may even reduce it.}

An important observation in fermionic semantics is that when individually
optimizing the parameter of a new operator $\hat{\tau}_{i}$ in the
ansatz, we only need to calculate this parameter's gradient and do
not have to study the energy of the full hamiltonian $\hat{H}$. Here,
``full'' underscores that the Hamiltonian
includes all non-zero excitation terms, rendering a size of $O(N^{4})$
\cite{Quantum_computational_chemistry}. In this case, only those
terms in $\hat{H}$ that share indices with $\hat{\tau}_{i}$
contribute, i.e., $\{p,q,r,s\}\bigcap\{a,b,i,j\}\neq\emptyset$. Here,
these two sets denote the indices of the Hamiltonian term and $\hat{\tau}_{i}$,
respectively. Thus, we only need to collect these relevant terms from
$\hat{H}$ to form a new sub-Hamiltonian $\hat{H}_{i}$ for computing
the gradient of $\hat{\tau}_{i}$. For a single excitation operator,
the size of the sub-Hamiltonian $\hat{H}_{i}$ scales as $O\left(\binom{N}{1}^{2}\!\!\!+\!\binom{N}{1}^{4}
\!\!\!-\!\binom{N-2}{1}^{2}\!\!\!-\!\binom{N-2}{1}^{4}\right)\!=\!O(N^{3})$.
Similarly, for a double excitation operator, this count also scales
as $O\left(\binom{N}{1}^{2}\!\!\!+\!\binom{N}{1}^{4}\!\!\!-\!\binom{N-4}{1}^{2}\!\!\!-\!\binom{N-4}{1}^{4}\right)\!=\!O(N^{3})$.

Employing the parameter-shift rule \cite{parameter_shift_rule_0,Parameter_shift},
the gradient of a variational parameter can be calculated by the discrepancy
in interested Hamiltonian expectation values between several quantum
circuits with shifted parameters. Using the Broyden-Fletcher-Goldfarb-Shanno
(BFGS) minimizer \cite{bfgs}, optimizing an ansatz $U(\boldsymbol{\theta}^{(m)})$
with $m$ variational parameters generally requires $O(m^{2})$ VQE
gradient evaluations \cite{ADAPT_VQE_QEB}. Since only one variational
parameter is involved in local VQE optimization, the number of gradient
evaluations required for convergence is $O(1)$. Given that the size
of the sub-Hamiltonian is $O(N^{3})$, the total measurement cost
for obtaining the converged parameter values of all excitation operators
in the pool (with a size of $O\!\left(\!\binom{N}{2}\!+\!\binom{N}{4}\!\right)\!\!=\!\!\!O(N^{4}))$
is $O(N^{7})$, which matches the selection cost of gradient-based
ADAPT-VQE. However, by adopting the optimized new parameter values, 
the starting point for the global VQE optimization in Step 5 can be 
positioned closer to the optimal solution compared to the warm-start 
strategy used in ADAPT-VQE, which initiates global VQE optimization 
with the new parameter set to 0. We term this approach ``hot-start''.
This may significantly reduce the number of global optimization iterations
and thus lower the total measurement cost. 
For strongly correlated molecules with larger sizes, the ansatz typically
contains more excitation operators and variational parameters. Hot-start
may introduce more pronounced benefits compared to warm-start in these
cases. This claim is demonstrated by the numerical experiments in
the next section.

The iteration termination condition for Param-ADAPT-VQE is that the 
parameter magnitude $|\theta_{i}^{*}|$ of the new operator falls below
a threshold $\epsilon>0$, or that the number of iterations exceeds
a limit $k_{\text{max}}>0$. 

We note that compared to ADAPT-VQE, Param-ADAPT-VQE modifies the criterion
for selecting excitation operators while retaining its fundamental
framework. 
It is therefore equally compatible with numerous derivative variants and inherits their respective advantages
--- such as adding multiple parallel excitation operators at once to reduce circuit depth \cite{ADAPT-VQE-TETRIS}, 
adopting advanced schemes for excitation operator implementation \cite{ADAPT_VQE_QEB,adapt_9_cnot}, 
and removing redundant operators that already existing in ansatz \cite{ADAPT_VQE_Pruned}.

\section{Numberical Results }\label{sec:Numberical-Results}

In this section, we present comprehensive numerical experiments to
validate the performance of the proposed Param-ADAPT-VQE algorithm
against the standard ADAPT-VQE. The analysis is divided into two dedicated
subsections with hierarchical research objectives: Subsection \ref{subsec:Detailed-Performance-Analysis}
focuses on a fine-grained, in-depth performance comparison using the
\ce{BeH2} molecule as a prototype system, demonstrating
that the gradient-based selection criterion in ADAPT-VQE is not infallible
and that the parameter constitutes a more robust criterion for
avoiding redundant operators. Subsection \ref{subsec:Benchmark-Studies}
further conducts systematic benchmark studies on three additional
molecular systems (\ce{LiH}, \ce{H2O} and \ce{NH3}), with a focus on three key practical
metrics --- energy error, parameter count, and measurement cost ---
to verify the scalability and effectiveness of Param-ADAPT-VQE across
diverse molecular systems. Collectively, these results provide both
microscopic mechanistic insights and macroscopic practical validation
for the proposed algorithm's improvements over the ADAPT-VQE.

In this work, electronic integrals and molecular point groups are
computed using PySCF \cite{pyscf}, whereas second quantization, Jordan-Wigner
transformation, and quantum circuit simulations are carried out via
MindSpore Quantum \cite{MindQuantum}. Excitation operators are implemented
based on the scheme \cite{efficient_excitations_circ,ADAPT_VQE_QEB}.
All computational procedures adopt the STO-3G basis set with no frozen
orbitals considered, and the BFGS algorithm \cite{bfgs} integrated
within the SciPy Python library \cite{scipy} is applied to all optimization
processes. The excitation operator pool is constructed based on the
Hamiltonian-Informed UCCSD ansatz \cite{HiUCCSD}, which leverages
information inherent in the molecular Hamiltonian to drastically reduce
the number of excitation operators compared to UCCSD, without compromising
its expressibility. All simulations in this work are performed under
idealized scenarios that exclude both sampling noise and hardware-related
noise. 

\begin{figure}
\centering
\includegraphics[scale=0.53]{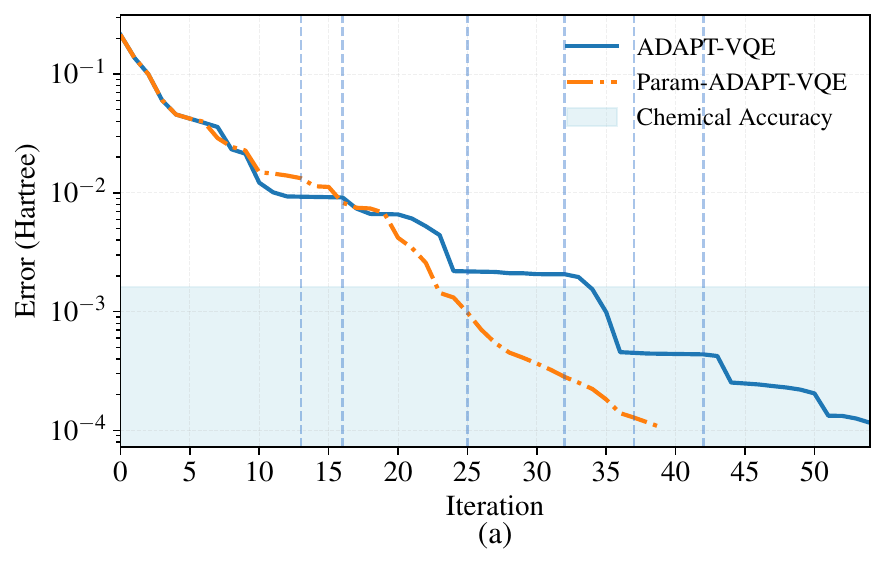}
\includegraphics[scale=0.53]{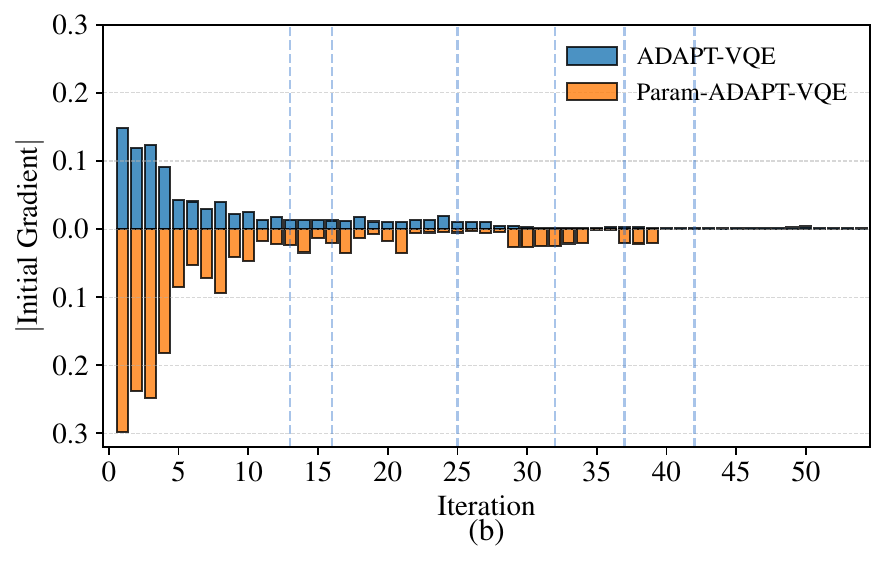}
\includegraphics[scale=0.53]{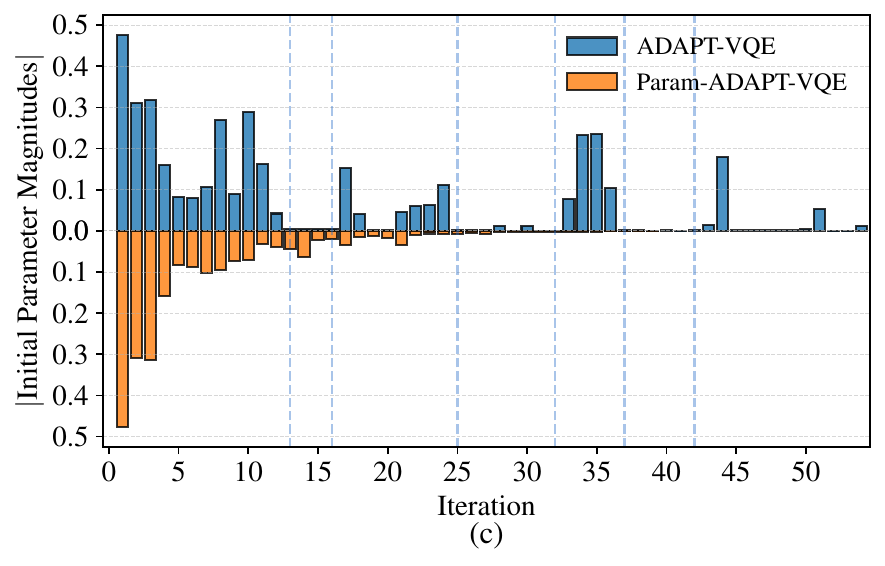}
\includegraphics[scale=0.53]{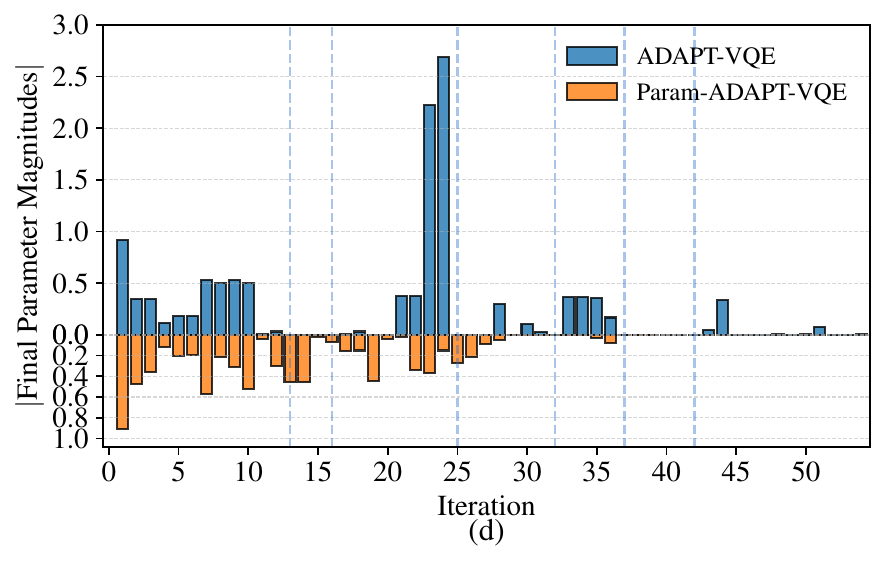}
\caption{Variations in (a) energy error, (b) initial gradient magnitudes of newly added operators, (c) initial parameter magnitudes of newly added operators, and (d) final parameter magnitudes of all variational parameters in the ansatz with respect to iteration number in the ground-state simulation of the \ce{BeH2} molecule (R(Be-H)=2.6\AA).
 }\label{fig:BeH2_sample}
\end{figure}

\subsection{Detailed Performance Analysis on \ce{BeH2} molecule}\label{subsec:Detailed-Performance-Analysis}

\begin{figure*}[!ht]
\centering
\includegraphics[scale=0.49]{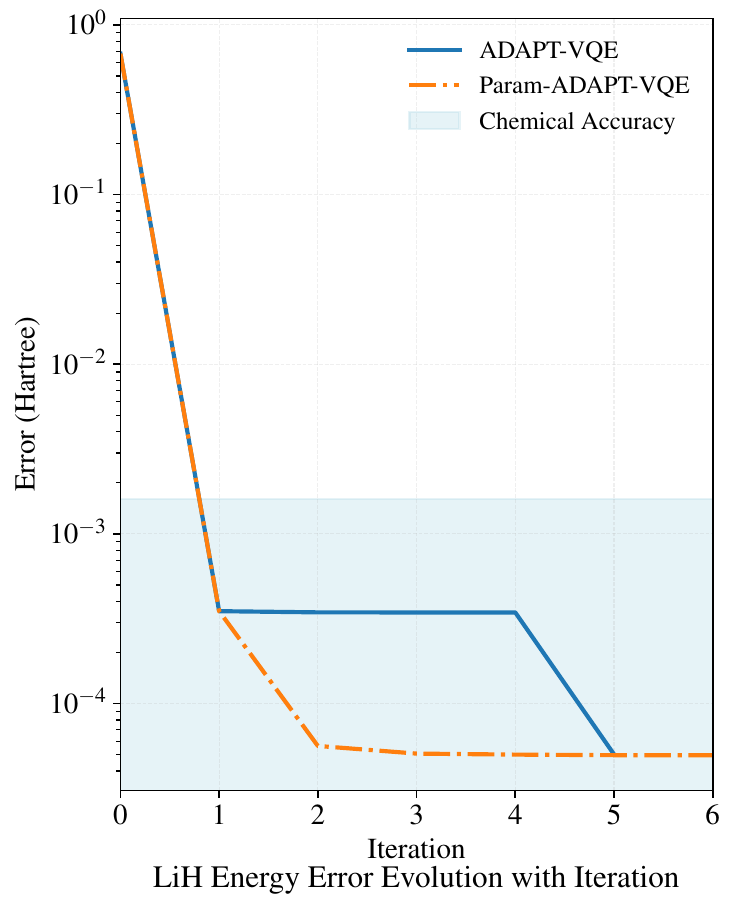}\includegraphics[scale=0.49]{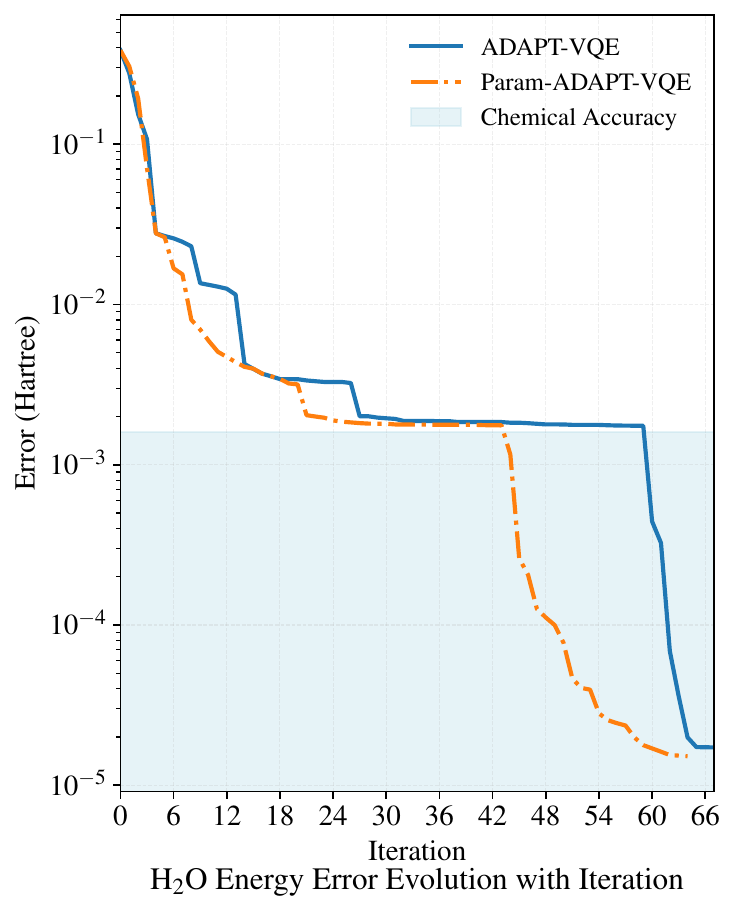}\includegraphics[scale=0.49]{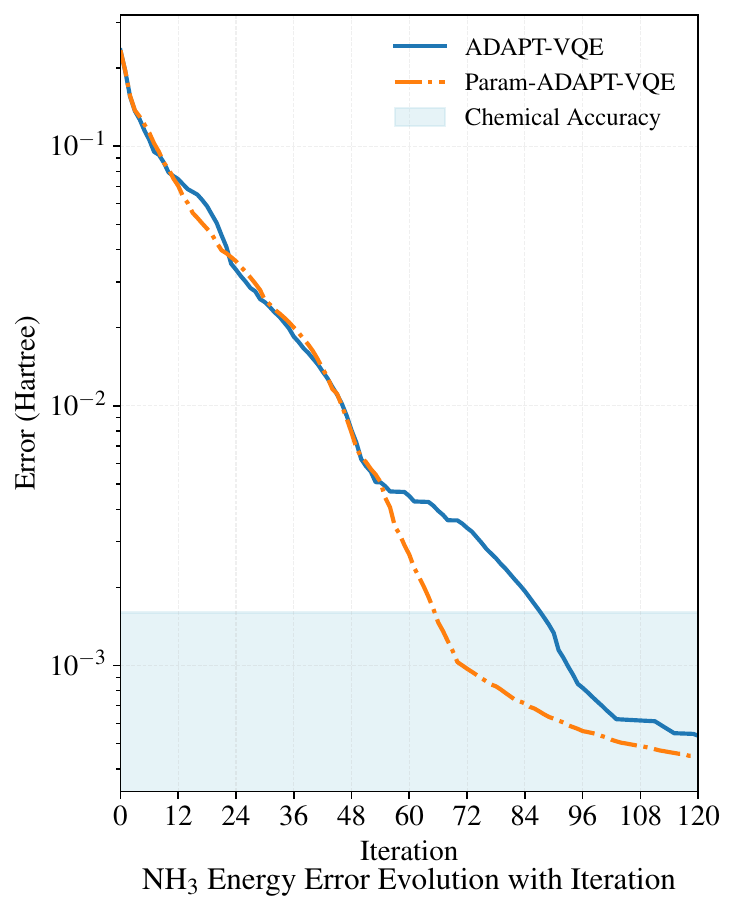}
\includegraphics[scale=0.49]{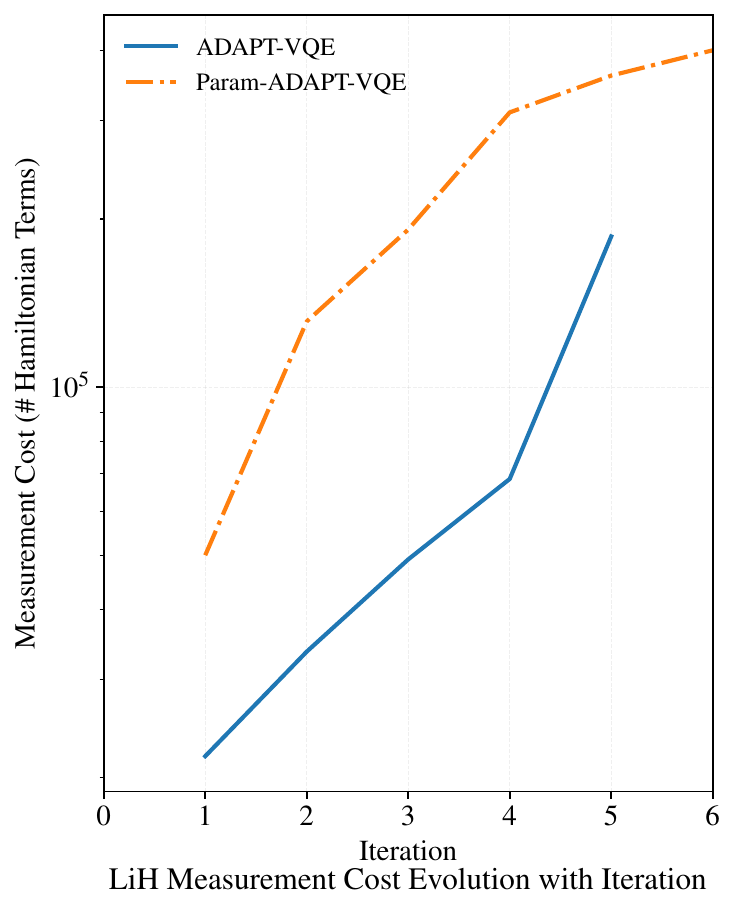}\includegraphics[scale=0.49]{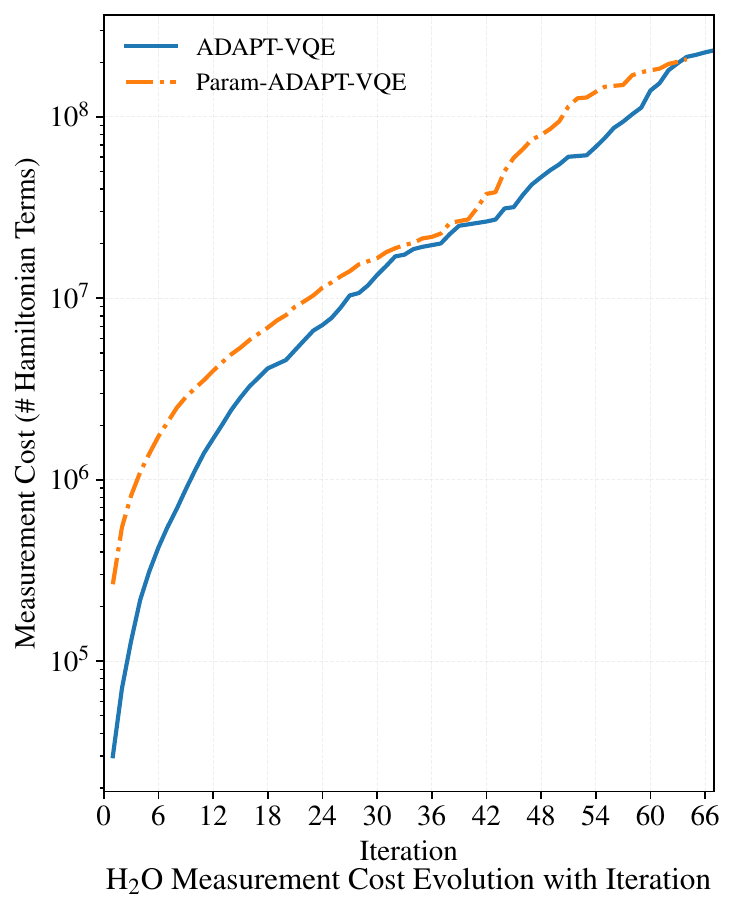}\includegraphics[scale=0.49]{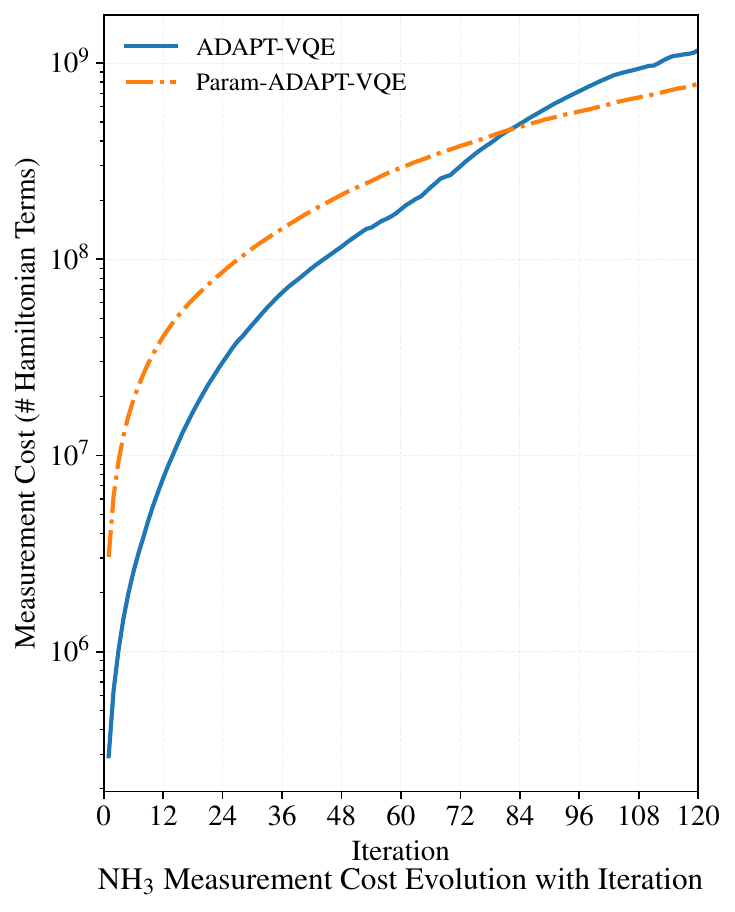}
\caption{(a--c) Energy error and (d--f) measurement cost as functions of
iteration number on \ce{LiH}, \ce{H2O},
and \ce{NH3} with uniformly stretched bond lengths of 3.24 \AA, 2.06 \AA, and 1.6 \AA, respectively. }\label{fig:2}
\end{figure*}

To ground this discussion in concrete context, we take the ground-state simulation of the \ce{BeH2} molecule with two 
symmetrically stretched bonds and R(Be-H)=2.6\AA \quad as a paradigmatic example.
Strongly correlated systems such as this are notably challenging, typically requiring long ansatze to reach sufficient computational accuracy. 
We note that while chemical accuracy (1.6 mHartree) is defined in the complete basis set limit or relative to experimental results, rather than the STO-3G basis set adopted in this work, it is nonetheless marked in the figures as a convention to facilitate readers in making comparisons with results from other related studies  \cite{ADAPT-VQE, ADAPT_VQE_Pruned, ADAPT_VQE_QEB, VQE_point_group_symmetry}.
In the discussion, we sometimes impose more stringent accuracy requirements, such as $10^{-4}$ and $10^{-5}$ Hartree.
Furthermore, we emphasize that both Param-ADAPT-VQE and ADAPT-VQE can achieve higher accuracy by setting more stringent stopping conditions. In this section, the norm thresholds for Param-ADAPT-VQE and ADAPT-VQE are set to $10^{-4}$ and $10^{-3}$, respectively, and the maximum number of iterations is fixed at 120.

Figure~\ref{fig:BeH2_sample}(a) depicts the evolution of energy error relative to the FCI result for ADAPT-VQE and Param-ADAPT-VQE as a function of iteration number. It can be observed that to reach a comparable error level of $10^{-4}$ Hartree, ADAPT-VQE requires 54 excitation operators, whereas Param-ADAPT-VQE only uses 39, representing a 28\% reduction in operator count.

Here, an excitation operator is regarded as redundant if it contributes almost no energy reduction and its parameter value nearly vanishes after optimization \cite{ADAPT_VQE_Pruned}.
To identify the redundant operators involved in ansatz construction, the iterative process is partitioned into distinct regions using vertical blue dashed lines for the ADAPT-VQE results.
These regions alternate between error-decreasing intervals and flat regions: the former denotes stages where the newly added excitation operators effectively induce a reduction in energy error, while the latter corresponds to periods where the total energy remains nearly constant despite continued iteration and the introduction of additional operators.
Notably, ADAPT-VQE exhibits three distinct error plateaus —-- i.e., iteration steps from 13 to 16, 25 to 32, and 37 to 42 —-- demarcated by dashed lines. In contrast, Param-ADAPT-VQE achieves a consistent reduction in energy error at each iteration, with no redundant excitation operators being introduced.

Figure~\ref{fig:BeH2_sample}(b) shows the initial gradient magnitudes of newly added excitation operators at each iteration.
ADAPT-VQE exhibits considerable gradient magnitudes across both error-decreasing intervals and flat regions; however, only negligible performance improvement is observed in the latter according to Figure~\ref{fig:BeH2_sample}(a).
In contrast, for the iteration steps around the 25th, the excitation operators selected by Param-ADAPT-VQE exhibit smaller gradient magnitudes than those of ADAPT-VQE, yet induce a more pronounced reduction in energy error.
These observations imply that operators with the largest initial gradient magnitudes do not necessarily yield effective energy reduction, and that gradient  is not an infallible selection criterion.

Figure~\ref{fig:BeH2_sample}(c) presents the magnitudes of the parameter values of newly added excitation operators after VQE optimization at each iteration, which we refer to as the initial parameter magnitudes.
It can
be observed that in the error flat regions of ADAPT-VQE, excitation
operators selected by the gradient-based criterion exhibit
nearly zero parameter values --- indicating these operators exert no
significant effect and are thus redundant. In contrast, operators selected via the parameter-based criterion in Param-ADAPT-VQE exhibit notable parameter magnitudes, corresponding to more pronounced contributions to the system. This figure reveals that parameter values serve as a more reliable selection criterion than gradient magnitudes for ansatz construction.

Figure~\ref{fig:BeH2_sample}(d) depicts the final optimized parameter magnitudes of all variational parameters after the last iteration. It can be seen that within the flat regions, most of the parameter values for ADAPT-VQE are nearly zero, confirming these operators are redundant. In contrast, Param-ADAPT-VQE yields considerable magnitudes for most parameters, demonstrating their significant contributions to the system's total energy.

\subsection{Benchmark Studies on Extended Molecular Systems }\label{subsec:Benchmark-Studies}

This subsection presents systematic benchmark studies of Param-ADAPT-VQE
on three prototypical molecules (LiH, \ce{H2O}, \ce{NH3}),
evaluating its performance against ADAPT-VQE across three key practical
metrics: energy error, parameter count, and measurement cost.
To emulate strongly correlated systems, the bond lengths of these molecules are uniformly stretched, i.e., R(Li-H)=3.24\!\! \AA \! for LiH, R(O-H)=2.06\!\! \AA \! for \ce{H2O}, and R(N-H)=1.6\! \AA \! for \ce{NH3}, respectively.

Figure~\ref{fig:2}(a) - (c) depicts the evolution of energy error for the three molecules as a function of increasing iteration number (i.e., the number of excitation operators) for the two competing algorithms, respectively.
Overall, Param-ADAPT-VQE achieves higher accuracy with
fewer excitation operators compared to ADAPT-VQE. 

Specifically, Figure~\ref{fig:2}(a) shows that for \ce{LiH} to reach an accuracy of $10^{-4}$ Hartree,
ADAPT-VQE requires 5 excitation operators while Param-ADAPT-VQE only
needs 2, representing a 60.00\% reduction. 
Figure~\ref{fig:2}(b) illustrates that for \ce{H2O} at the $10^{-4}$ Hartree
accuracy, ADAPT-VQE demands 62 excitation operators versus merely
49 for Param-ADAPT-VQE, a 20.97\% reduction. 
Figure~\ref{fig:2}(c) demonstrates that to achieve $10^{-3}$ Hartree accuracy, ADAPT-VQE
needs 93 excitation operators whereas Param-ADAPT-VQE requires 72,
corresponding to a 22.58\% reduction.

Figure~\ref{fig:2}(d) - (f) presents the corresponding measurement cost as a function of iteration number for the three molecules.
The measurement cost here is quantified by the number of terms in the fermionic Hamiltonian
for which expectation values need to be considered. 
To provide a general reference, the simultaneous measurement \cite{VQE_simutameously_measurement_N3} and measurement reuse \cite{VQE_measurement_reuse} techniques 
are not considered in this work. We observe that the measurement cost
of Param-ADAPT-VQE rises at a faster rate in the initial stage, owing
to the additional local VQE optimization performed on all excitation
operators in the pool during its operator selection process. However,
as iterations proceed, the variational parameters in the ansatz accumulate
progressively, and the measurement cost required for each global optimization
step increases substantially, ultimately surpassing that of the local
VQE optimization. 

Specifically, Figure~\ref{fig:2}(a) and (d) shows that for LiH at an energy error of $10^{-4}$ Hartree, the measurement cost of ADAPT-VQE reaches $1.86\times10^6$, while that of Param-ADAPT-VQE is only $1.31\times10^6$, representing a 29.55\% reduction. Figure~\ref{fig:2}(b) and (e) reveals that for \ce{H2O} at the same $10^{-4}$ Hartree accuracy, the measurement cost of ADAPT-VQE reaches $1.81\times10^9$, while that of Param-ADAPT-VQE is only $8.57\times10^8$, corresponding to a 52.62\% reduction. Figure~\ref{fig:2}(c) and (f) demonstrates that for \ce{NH3} at an energy error of $10^{-3}$ Hartree, the measurement cost of ADAPT-VQE reaches $6.59\times10^9$, while that of Param-ADAPT-VQE is only $3.78\times10^9$, a 42.67\% reduction. A key insight emerging from these results is that Param-ADAPT-VQE surpasses ADAPT-VQE in both accuracy and quantum measurement cost for all systems after the 81st iteration.

In addition, performance comparisons between Param-ADAPT-VQE and ADAPT-VQE across an expanded set of molecules (\ce{H4}, \ce{LiH}, \ce{HF}, \ce{BeH2}, \ce{H2O}, and \ce{NH3} at varying bond lengths) are provided in the Appendix for reference. The key conclusions drawn from these extended analyses are generally consistent with the aforementioned findings.

The above results demonstrate that within the ADAPT-VQE framework, the parameter-based criterion for excitation operator selection effectively reduces the redundancy rate in the ansatz compared with the gradient-based criterion. Furthermore, the adoption of sub-Hamiltonian and hot-start yields a further reduction in quantum measurement cost relative to warm-start. In summary, Param-ADAPT-VQE exhibits distinct advantages over ADAPT-VQE with respect to both ansatz depth and  measurement cost.

\section{Conclusion}\label{sec:Conclusion}

In this work, we propose the Param-ADAPT-VQE, a novel algorithm for selecting
excitation operators via a parameter-based criterion. We characterize
the features of redundant operators through concrete case studies
and demonstrate that Param-ADAPT-VQE can effectively reduce the proportion
of redundant operators in the ansatz. A sub-Hamiltonian technique
is devised to avoid the significant increase in measurement cost incurred
by operator pool scanning. Furthermore, we adopt a hot-start optimization
strategy, which cut down the measurement cost associated with global
VQE optimization by place the global VQE optimization to a start point
closer to the optimal solution, therefore reduces the number of optimization
iterations. Numerical implementations on molecular systems including
LiH, \ce{H2O} and \ce{NH3} show that the Param-ADAPT-VQE
outperforms ADAPT-VQE in terms of the number of excitation operators,
computational accuracy, and measurement cost metrics. 
Moreover, our algorithm is fully compatible with modified versions
of ADAPT-VQE, thus allowing further specific performance gains.

\section*{Data availability}
All data were generated with the code publicly available at \href{https://atomgit.com/mindspore/mindquantum/tree/research/paper_with_code/Param_ADAPT_VQE}{https://atomgit.com/mindspore/mindquantum/} \href{https://atomgit.com/mindspore/mindquantum/tree/research/paper_with_code/Param_ADAPT_VQE}{tree/research/paper\_with\_code/Param\_ADAPT\_VQE}.

\section*{Code availability}
The code used for the numerical simulations has been made  publicly
available at
\href{https://atomgit.com/mindspore/mindquantum/tree/research/paper_with_code/Param_ADAPT_VQE}{https://atomgit.com/mind-} \href{https://atomgit.com/mindspore/mindquantum/tree/research/paper_with_code/Param_ADAPT_VQE}{spore/mindquantum/tree/research/paper\_with\_code/Pa-}\href{https://atomgit.com/mindspore/mindquantum/tree/research/paper_with_code/Param_ADAPT_VQE}{ram\_ADAPT\_VQE}.

\section*{Acknowledgements}
This work was supported by the Innovation Program for Quantum Science and Technology under Grant No. 2024ZD0300502, and the Beijing Nova Program under Grant No. 20240484652.

\appendix 
\section{}\label{appendix}
\begin{figure*}
	\includegraphics[scale=0.3]{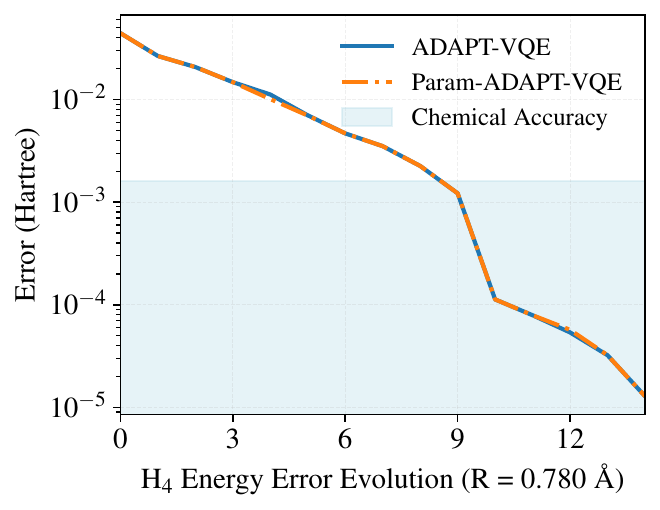}
	\includegraphics[scale=0.3]{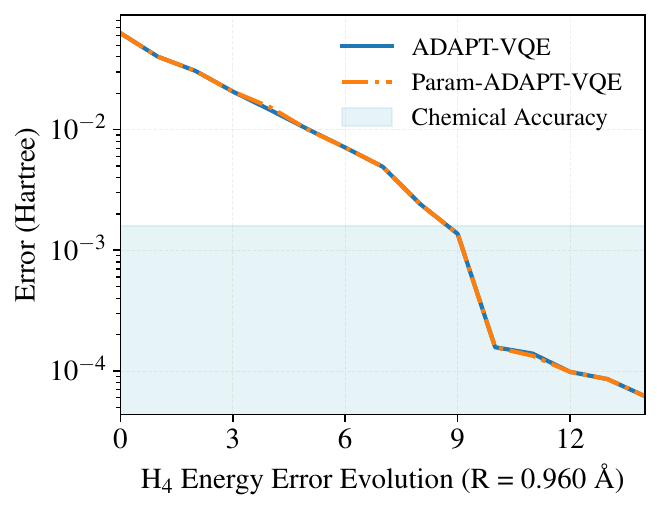}
	\includegraphics[scale=0.3]{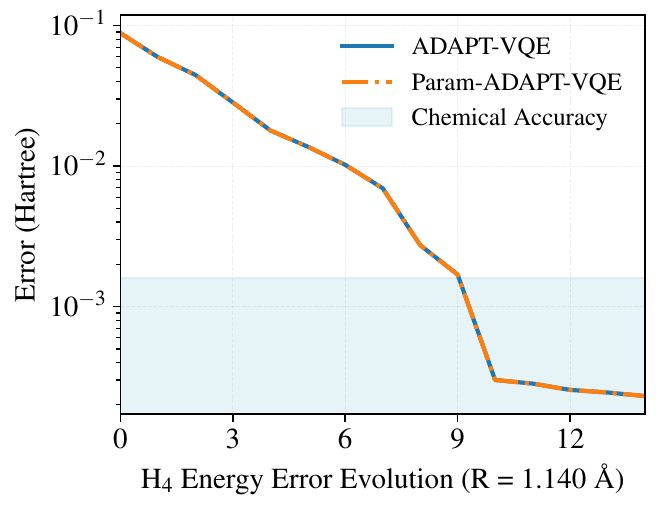}
	\includegraphics[scale=0.3]{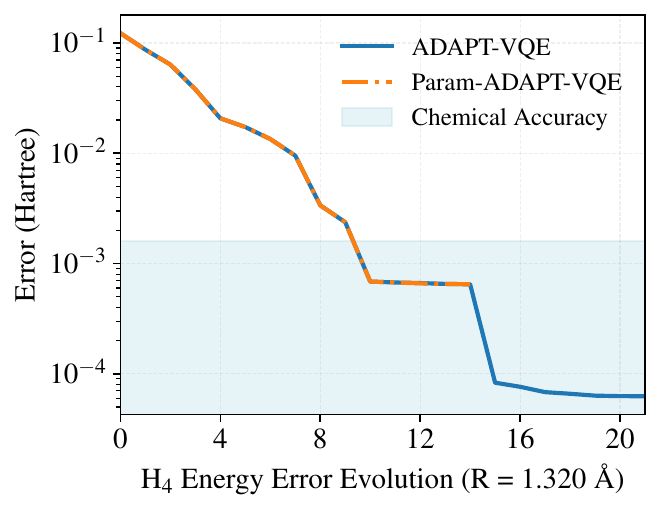}
	\includegraphics[scale=0.3]{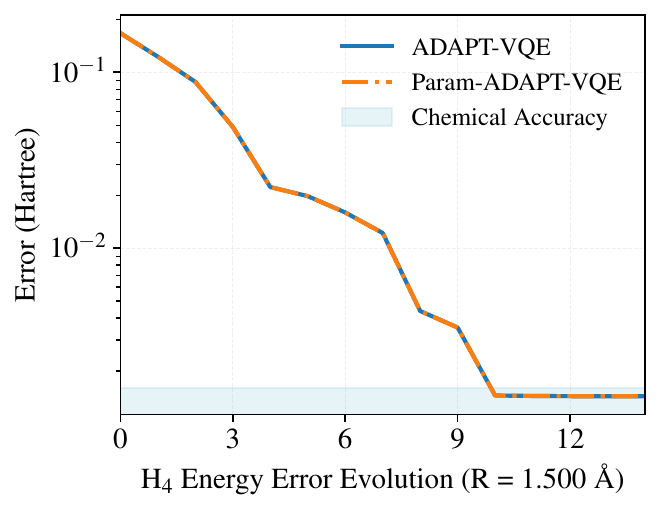}
	\includegraphics[scale=0.3]{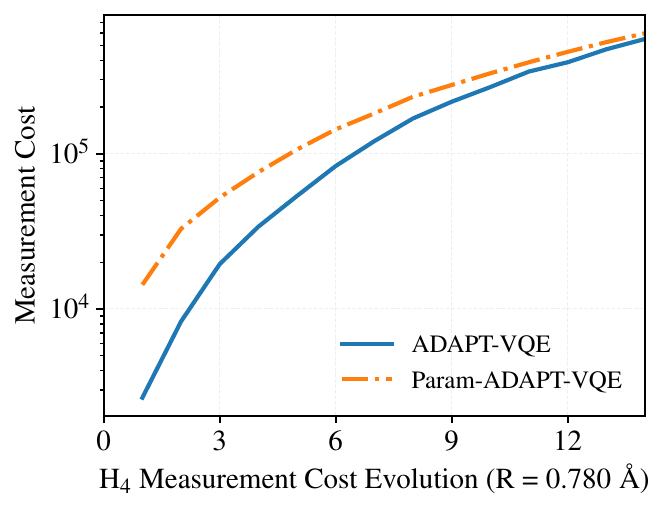}
	\includegraphics[scale=0.3]{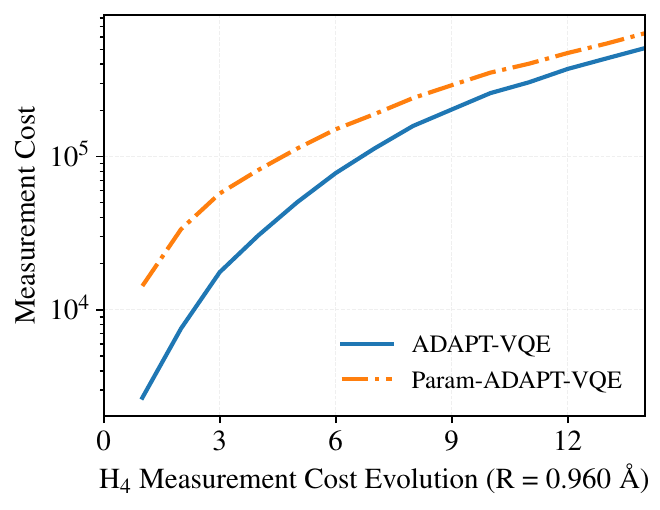}
	\includegraphics[scale=0.3]{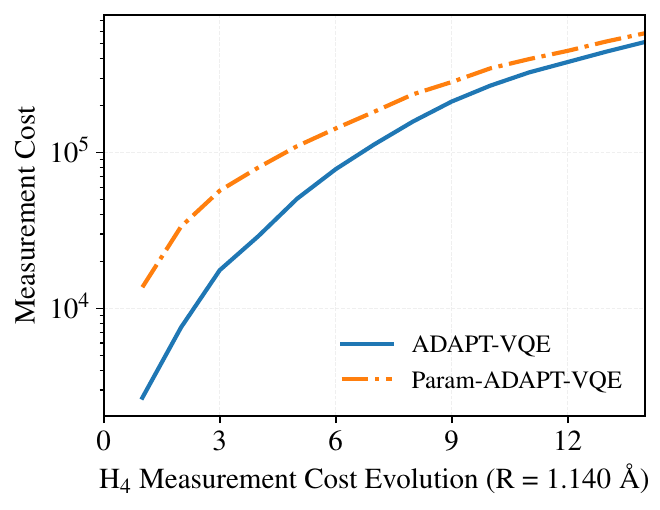}
	\includegraphics[scale=0.3]{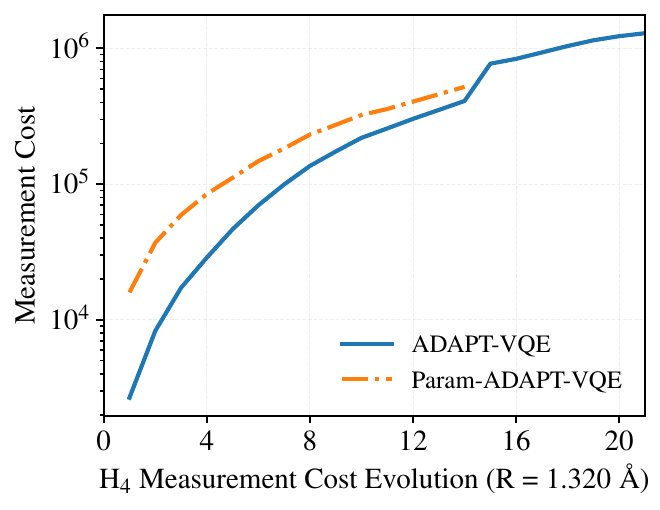}
	\includegraphics[scale=0.3]{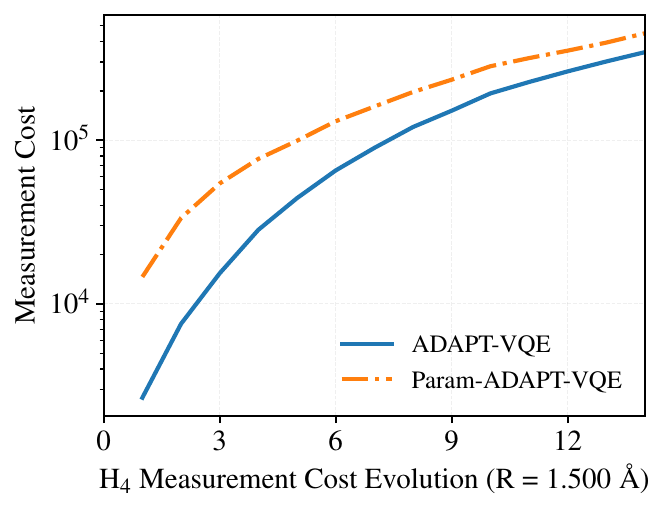}
	\includegraphics[scale=0.3]{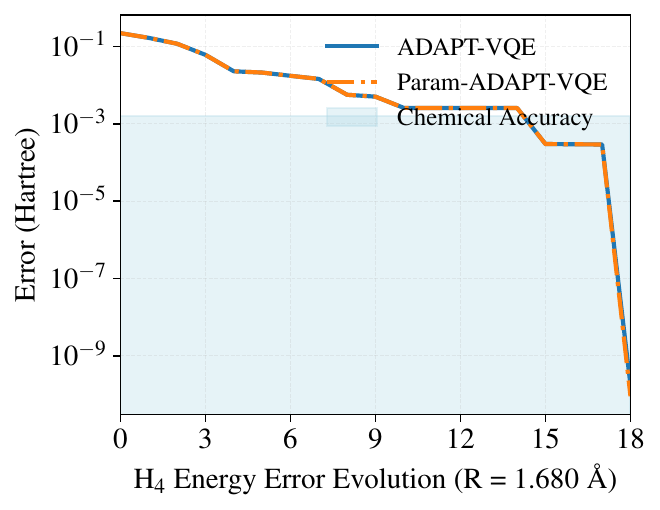}
	\includegraphics[scale=0.3]{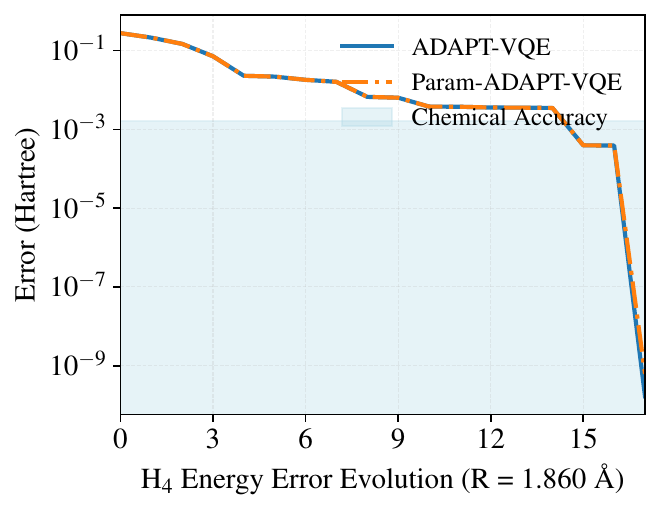}
	\includegraphics[scale=0.3]{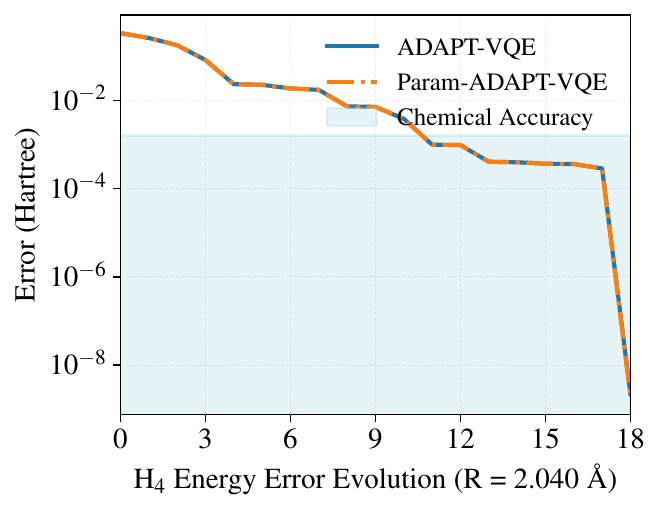}
	\includegraphics[scale=0.3]{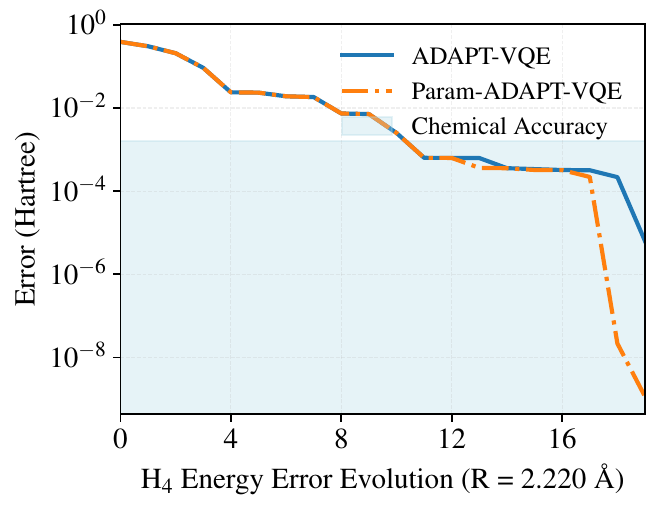}
	\includegraphics[scale=0.3]{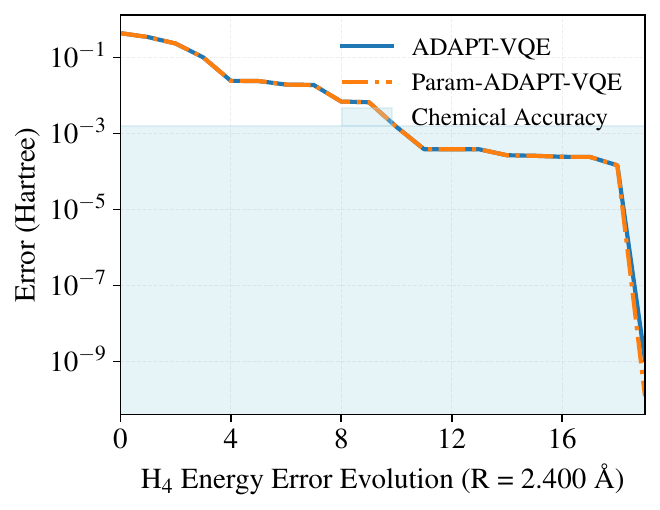}
	\includegraphics[scale=0.3]{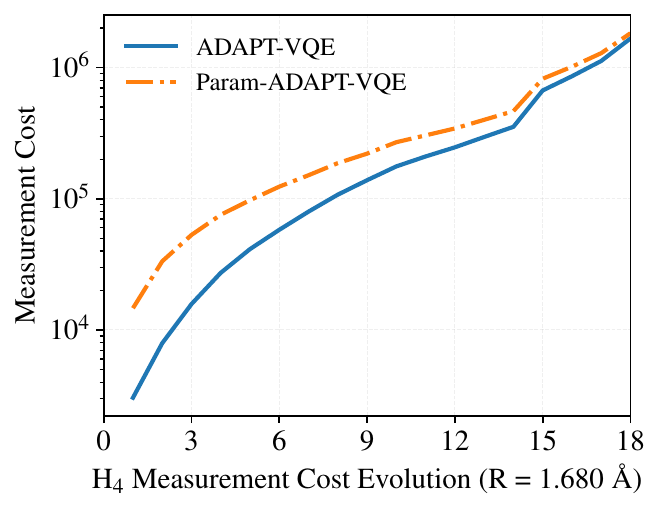}
	\includegraphics[scale=0.3]{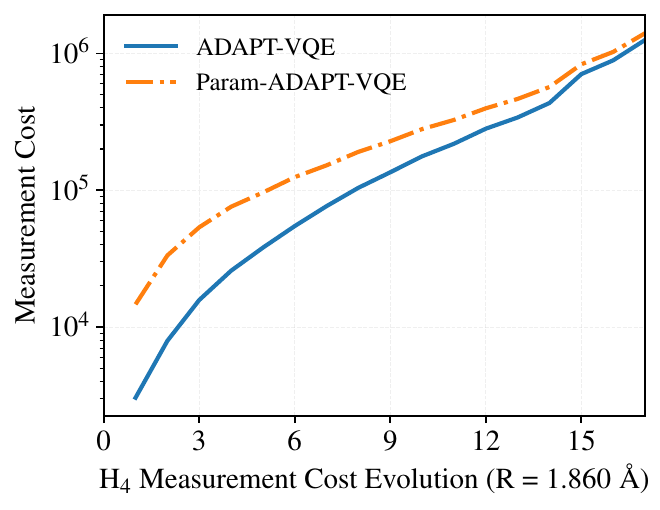}
	\includegraphics[scale=0.3]{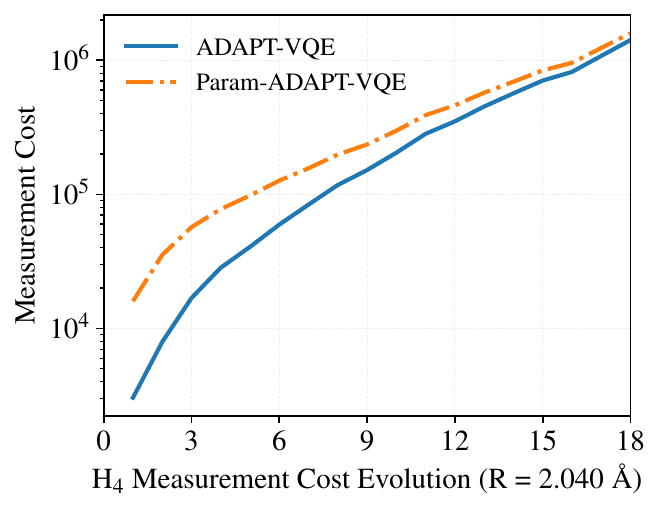}
	\includegraphics[scale=0.3]{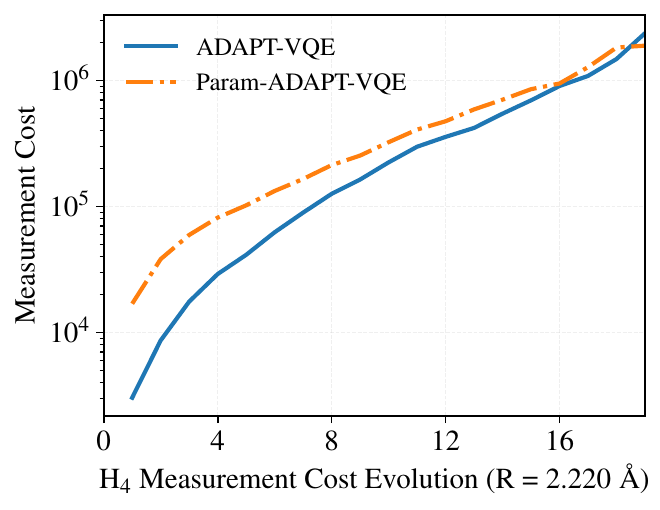}
	\includegraphics[scale=0.3]{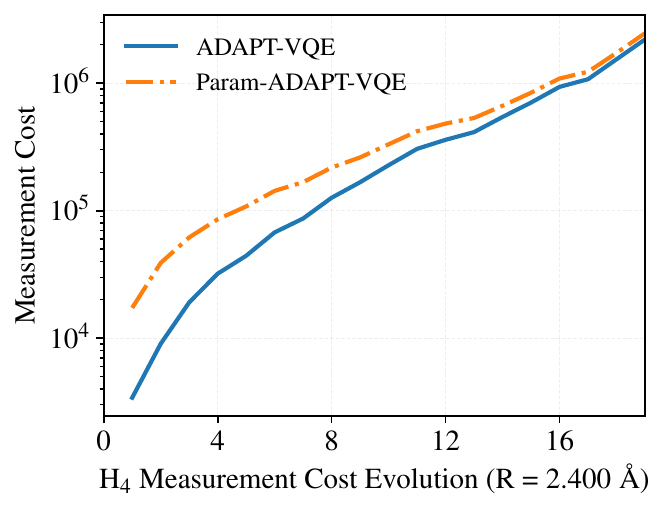}
\caption{Evolution of energy error and measurement cost with iteration for  \ce{H4}  at different bond lengths.\label{H4}
		}
\end{figure*}

\begin{figure*}
	\includegraphics[scale=0.3]{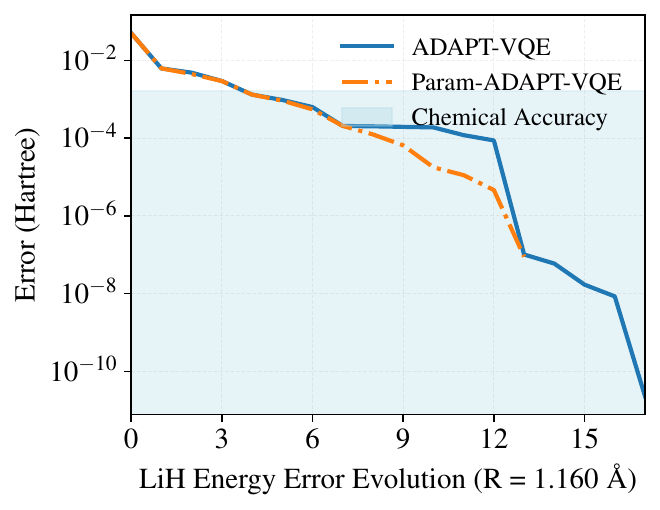}
	\includegraphics[scale=0.3]{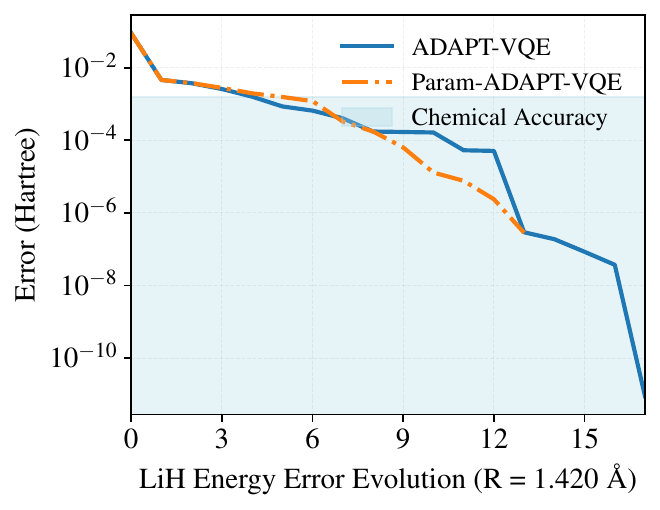}
	\includegraphics[scale=0.3]{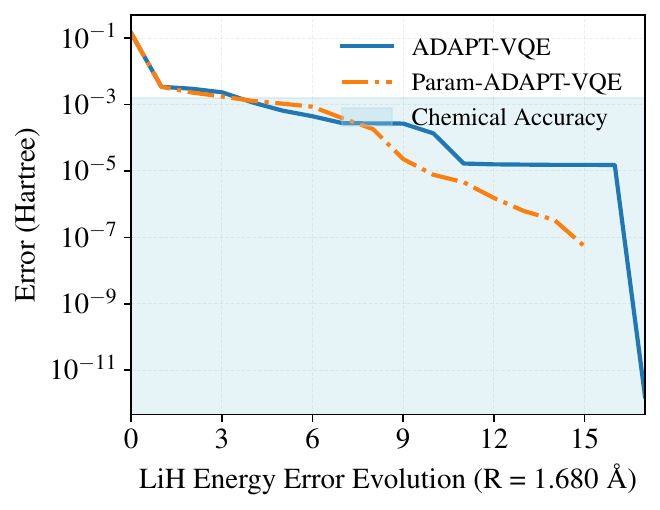}
	\includegraphics[scale=0.3]{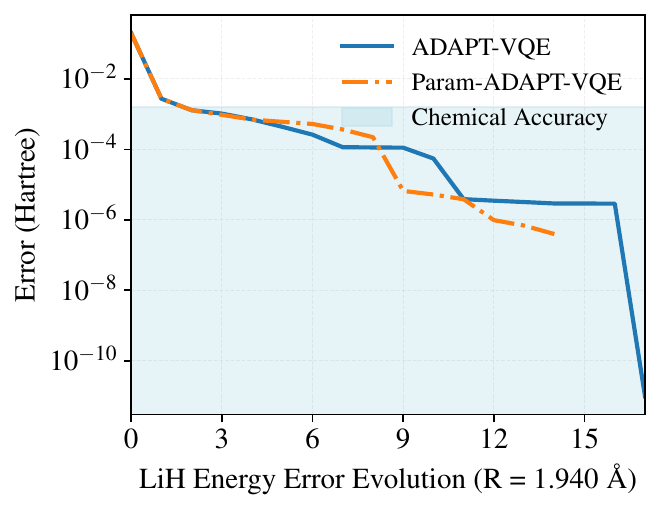}
	\includegraphics[scale=0.3]{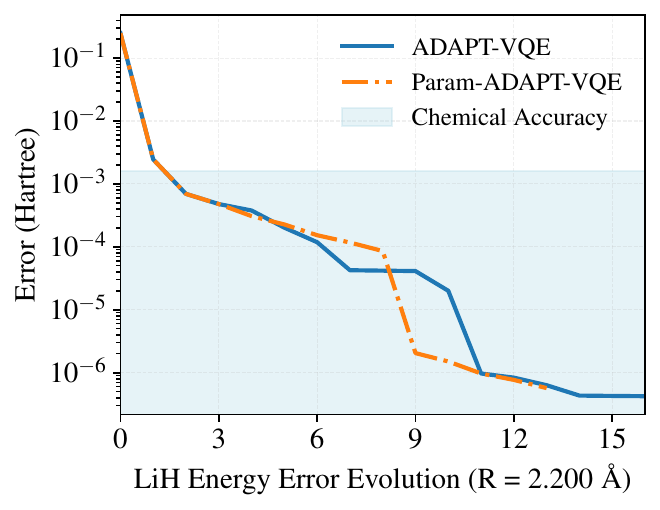}
	\includegraphics[scale=0.3]{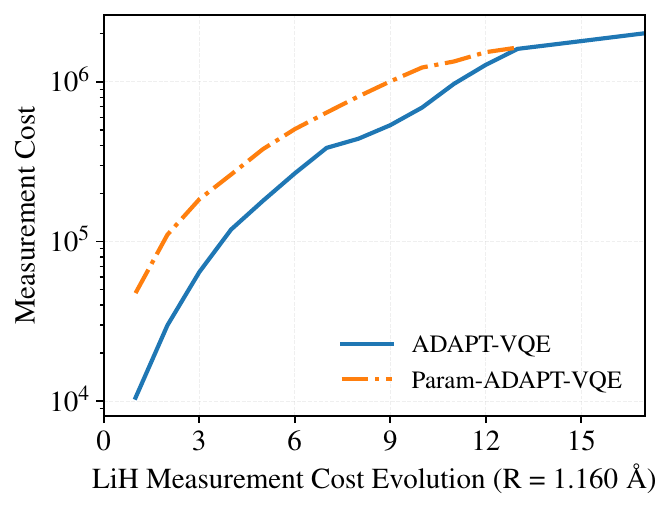}
	\includegraphics[scale=0.3]{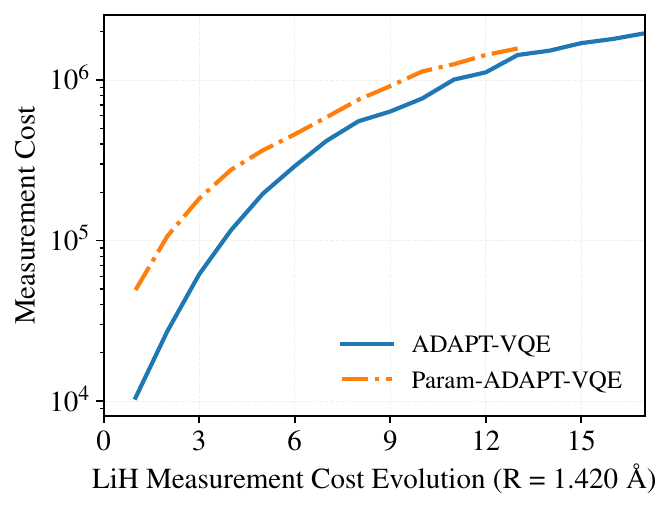}
	\includegraphics[scale=0.3]{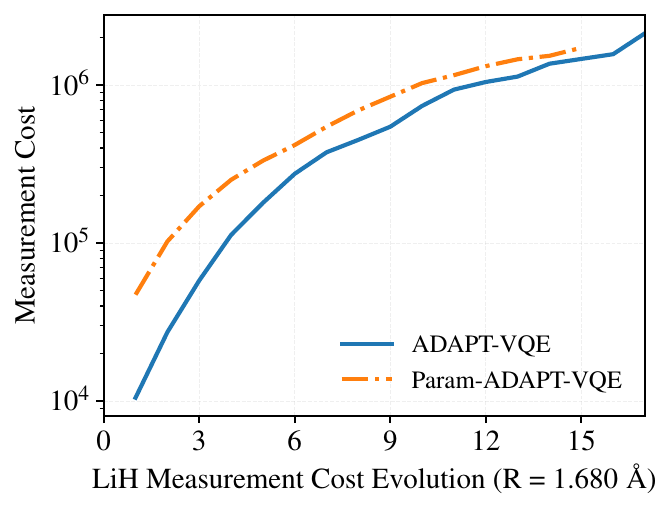}
	\includegraphics[scale=0.3]{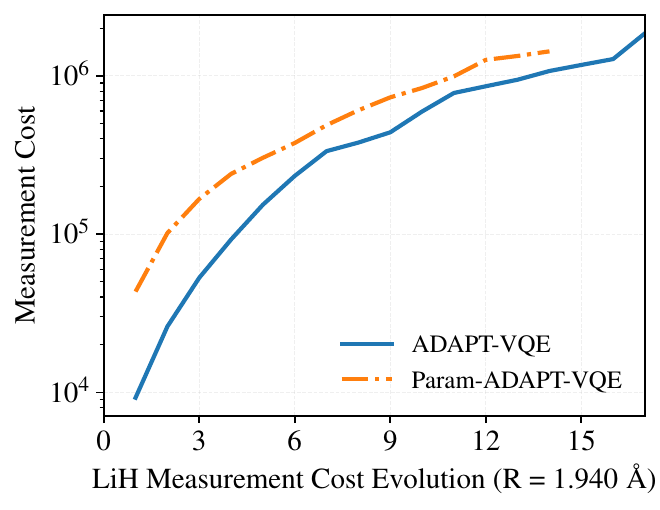}
	\includegraphics[scale=0.3]{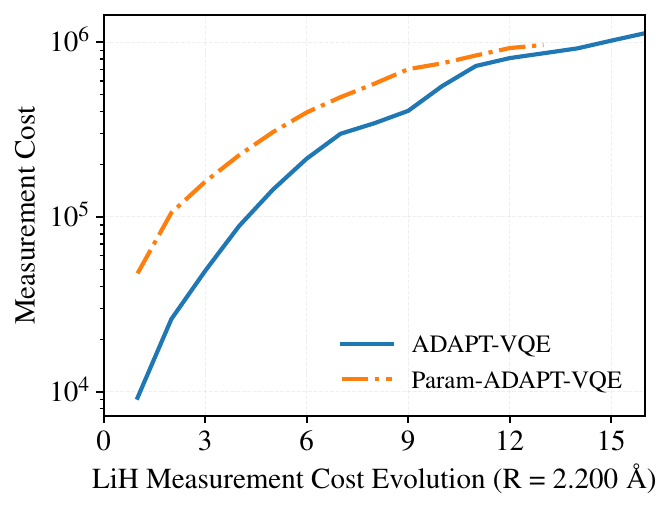}
	\includegraphics[scale=0.3]{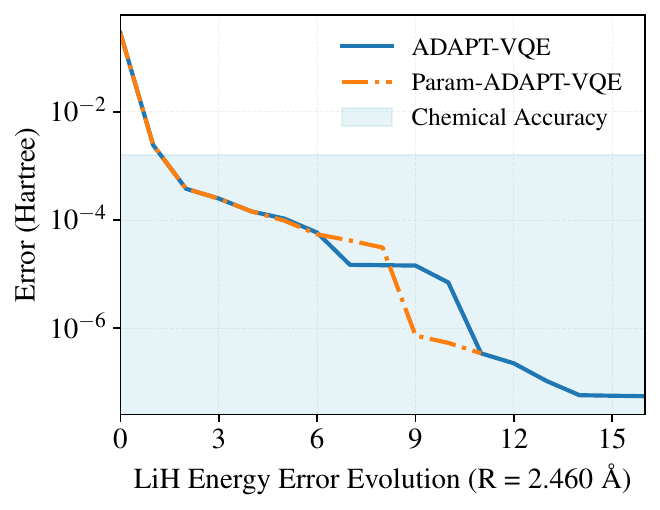}
	\includegraphics[scale=0.3]{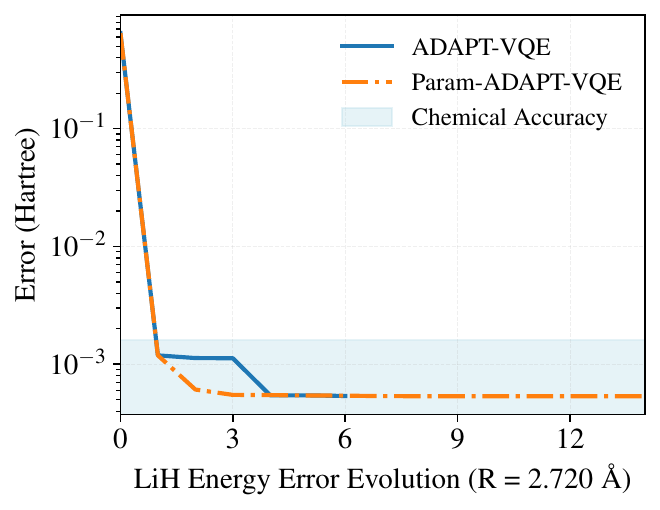}
	\includegraphics[scale=0.3]{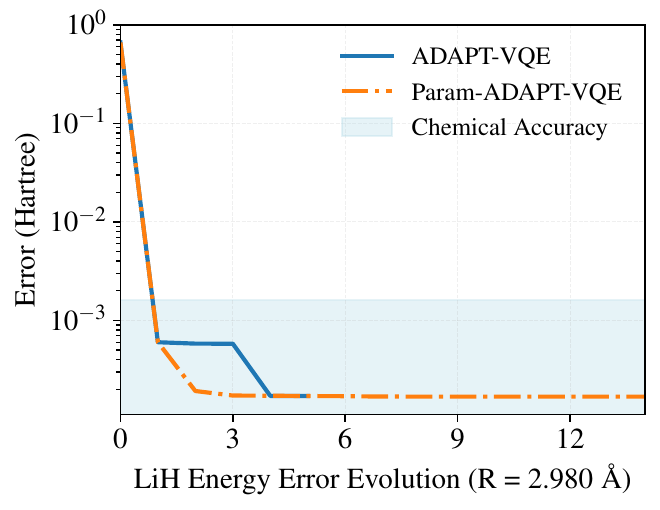}
	\includegraphics[scale=0.3]{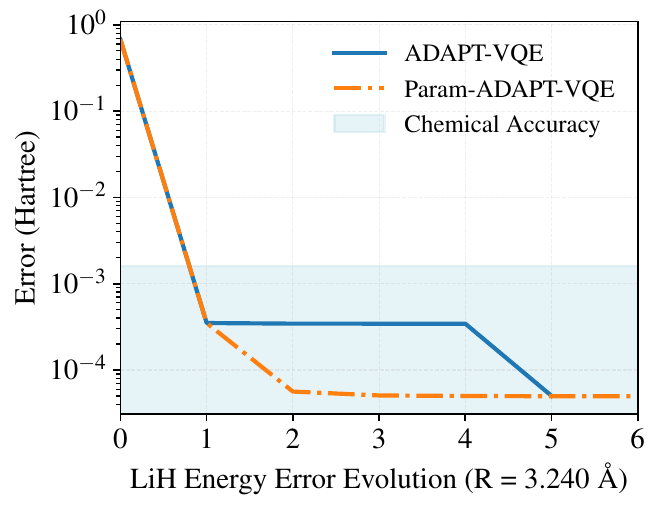}
	\includegraphics[scale=0.3]{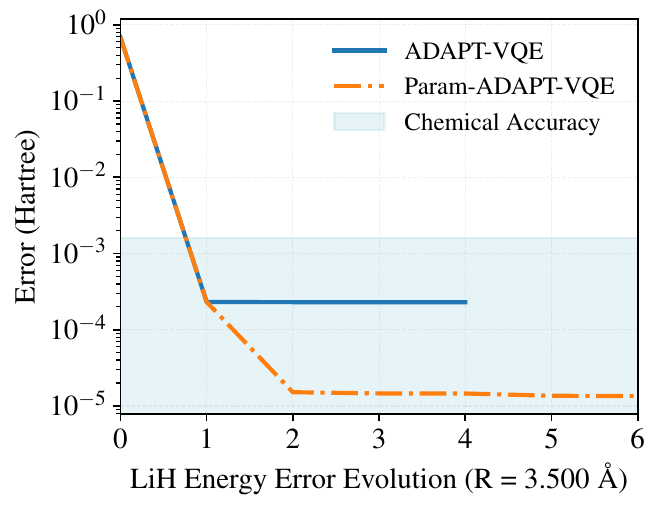}
	\includegraphics[scale=0.3]{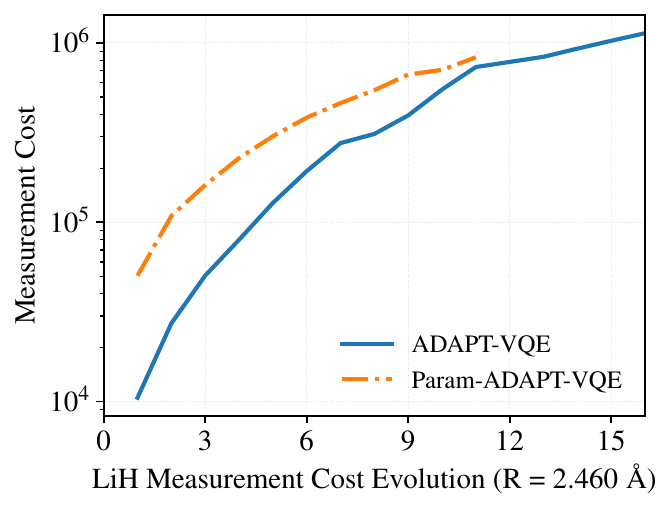}
	\includegraphics[scale=0.3]{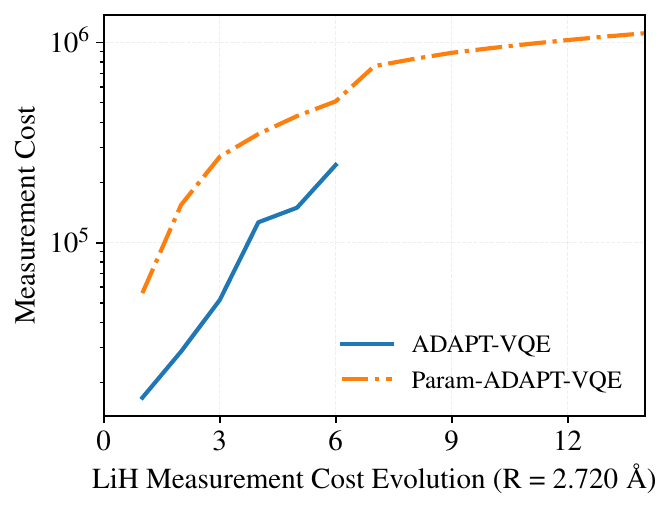}
	\includegraphics[scale=0.3]{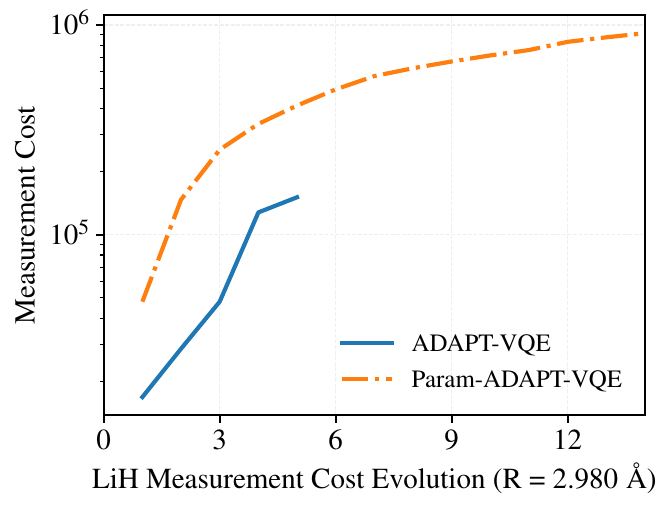}
	\includegraphics[scale=0.3]{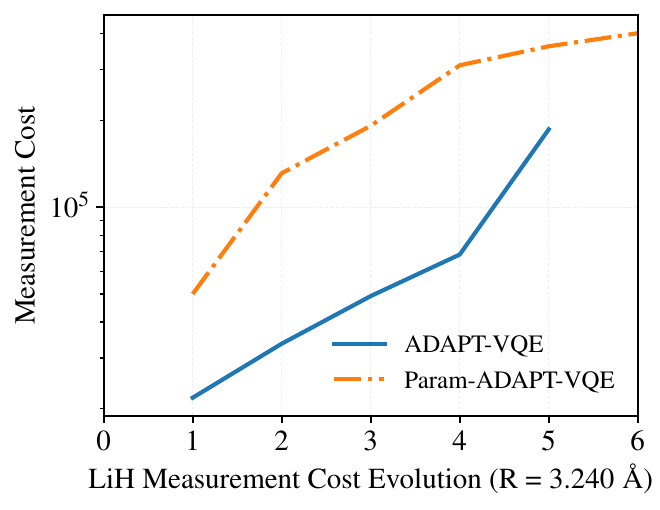}
	\includegraphics[scale=0.3]{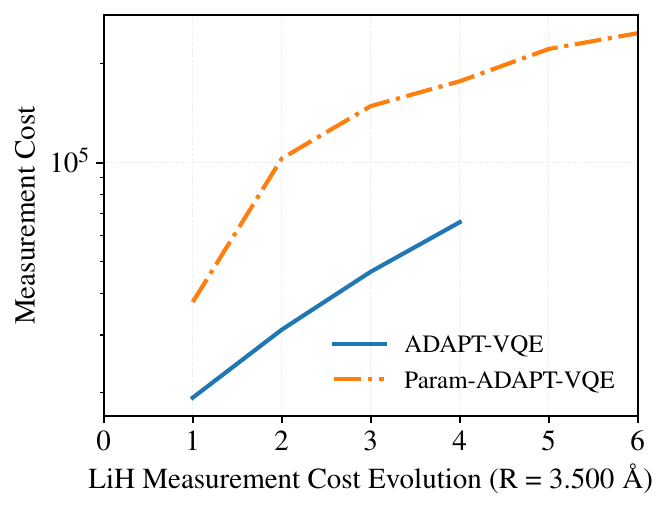}
	\caption{Evolution of energy error and measurement cost with iteration for \ce{LiH}  at different bond lengths.\label{LiH}
	}
\end{figure*}

\begin{figure*}
	\includegraphics[scale=0.3]{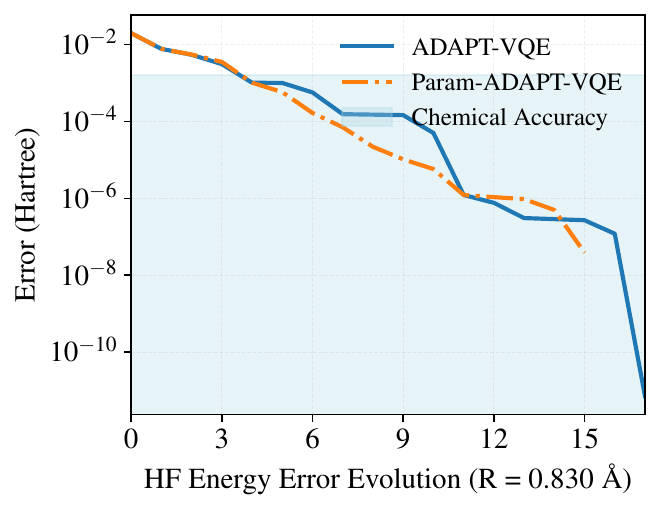}
	\includegraphics[scale=0.3]{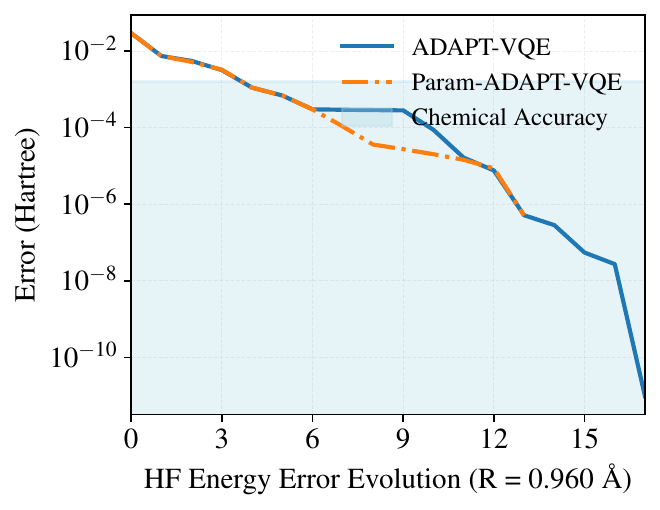}
	\includegraphics[scale=0.3]{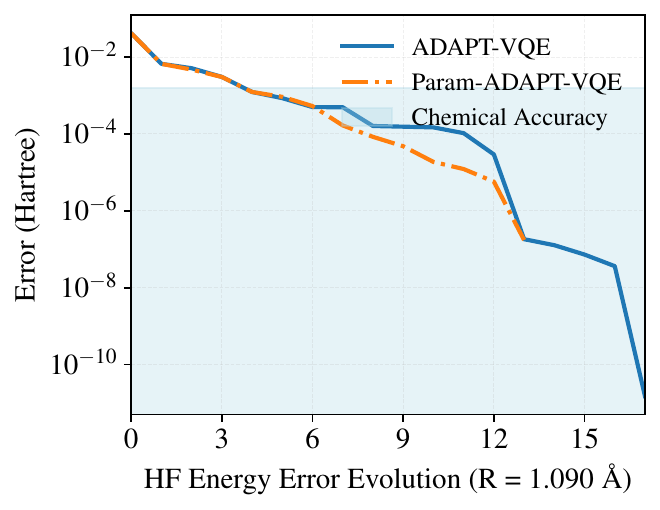}
	\includegraphics[scale=0.3]{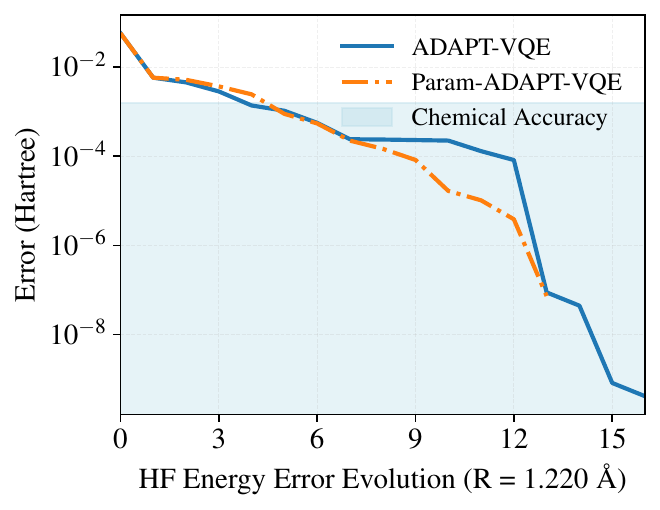}
	\includegraphics[scale=0.3]{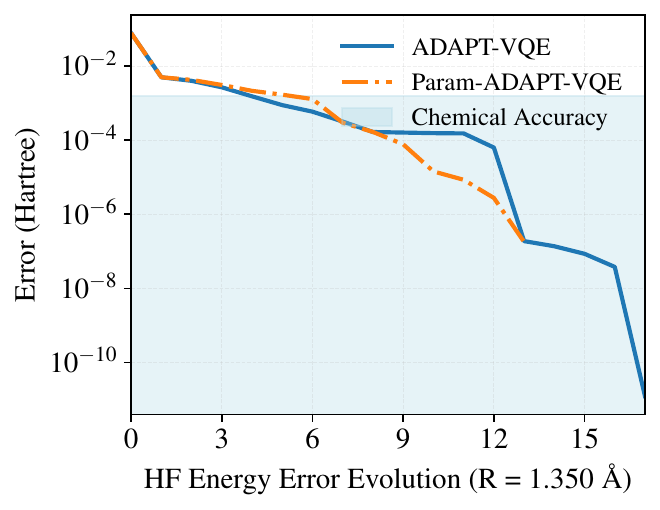}
	\includegraphics[scale=0.3]{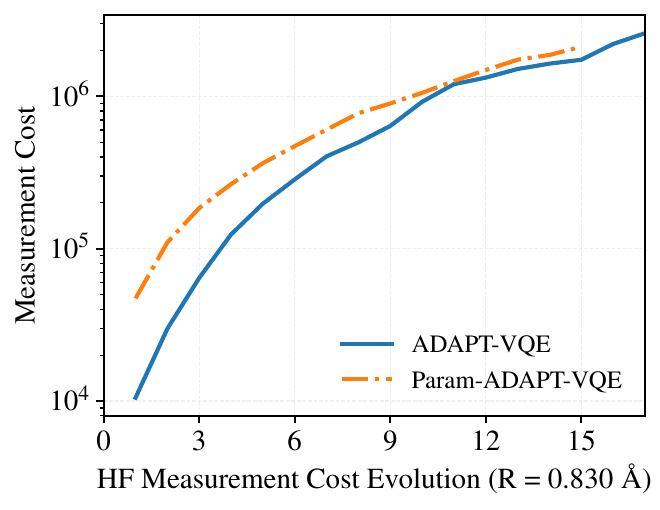}
	\includegraphics[scale=0.3]{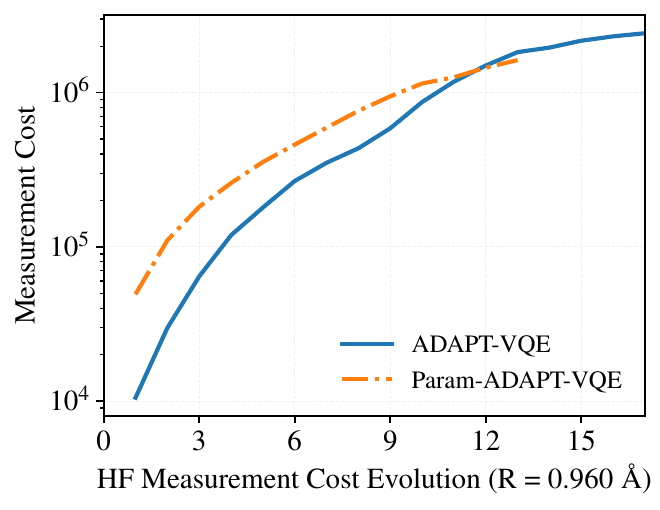}
	\includegraphics[scale=0.3]{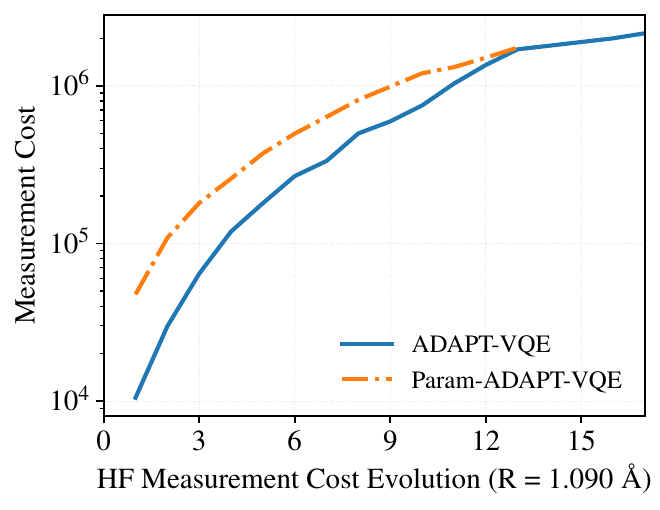}
	\includegraphics[scale=0.3]{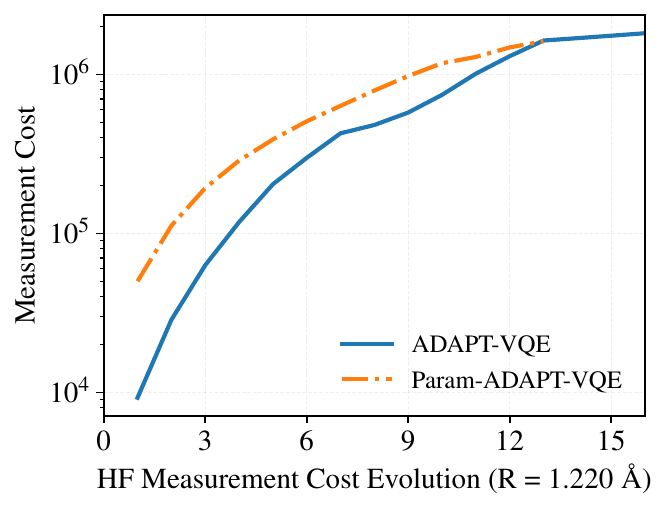}
	\includegraphics[scale=0.3]{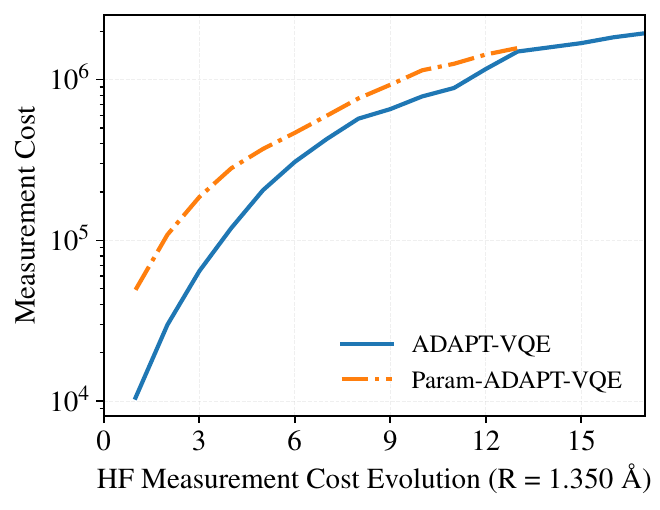}
	\includegraphics[scale=0.3]{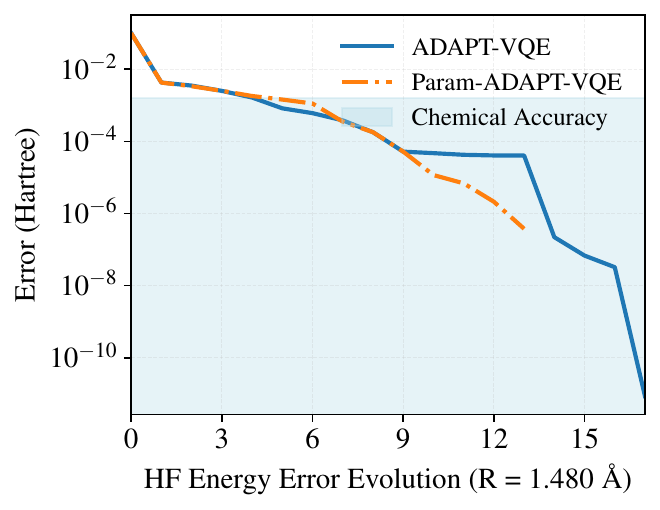}
	\includegraphics[scale=0.3]{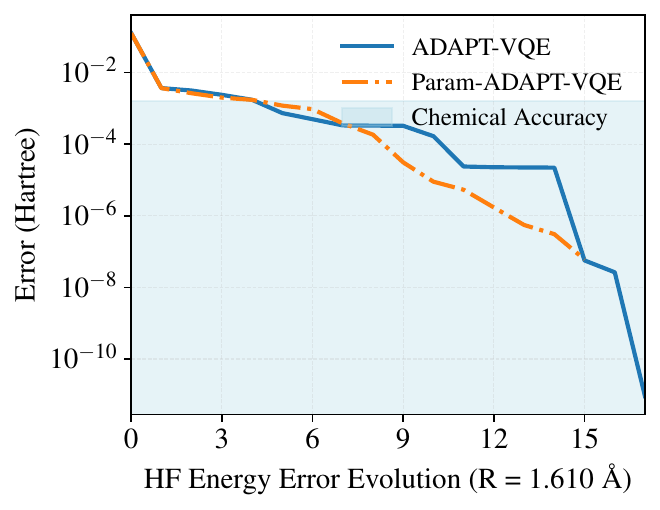}
	\includegraphics[scale=0.3]{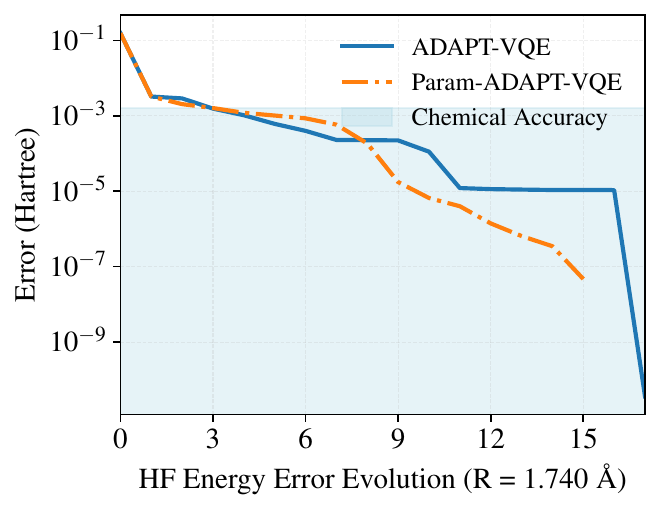}
	\includegraphics[scale=0.3]{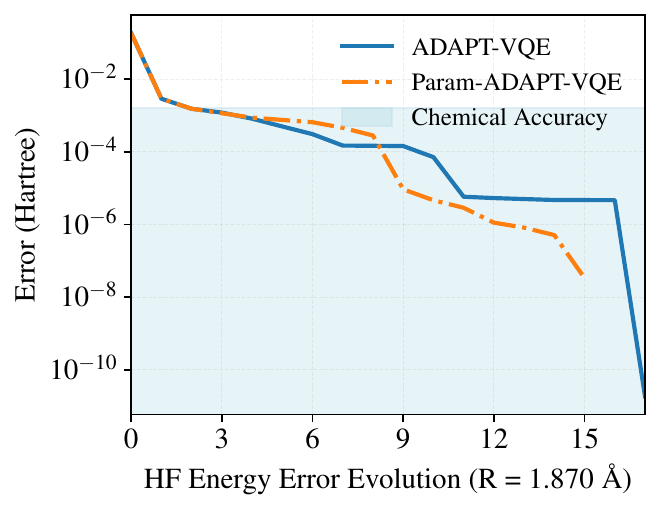}
	\includegraphics[scale=0.3]{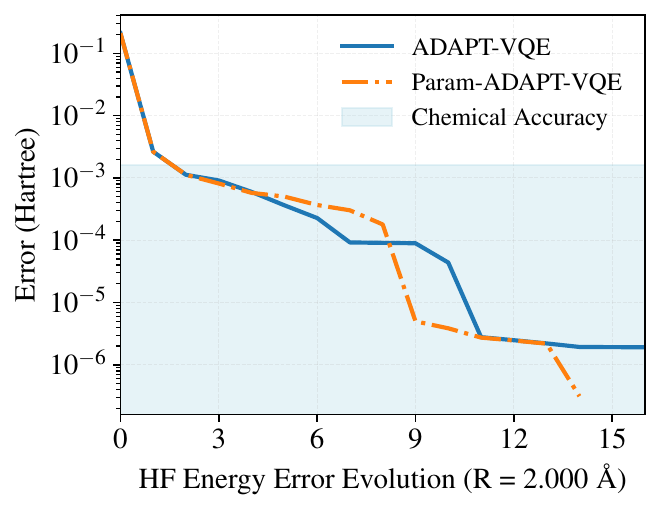}
	\includegraphics[scale=0.3]{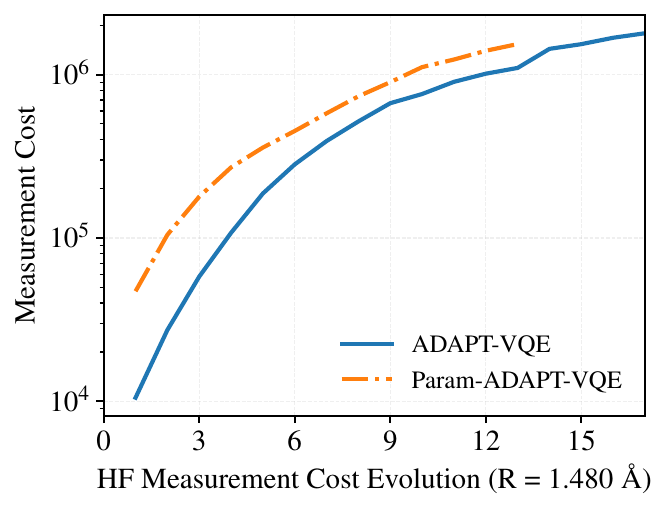}
	\includegraphics[scale=0.3]{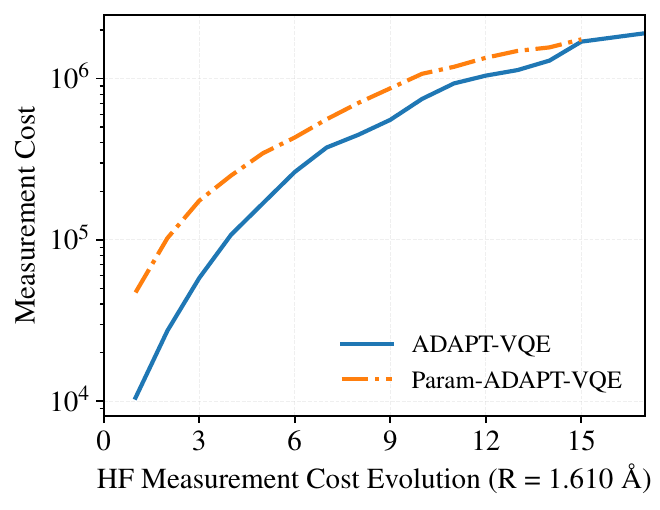}
	\includegraphics[scale=0.3]{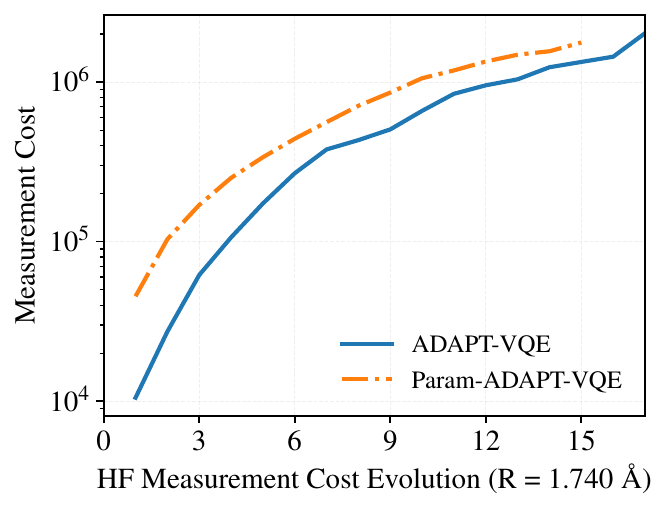}
	\includegraphics[scale=0.3]{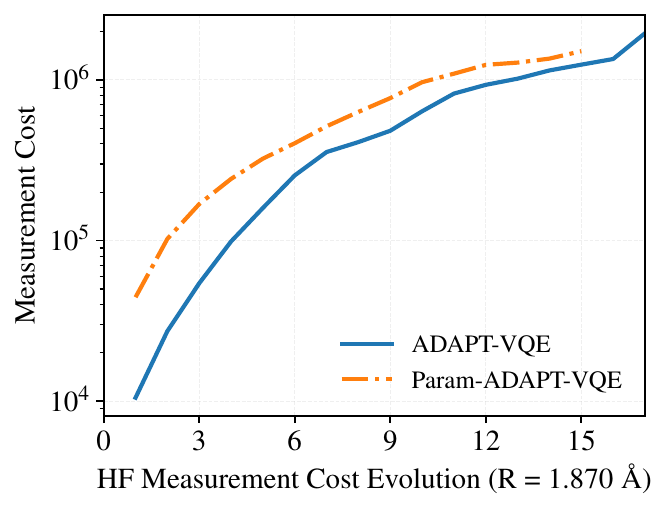}
	\includegraphics[scale=0.3]{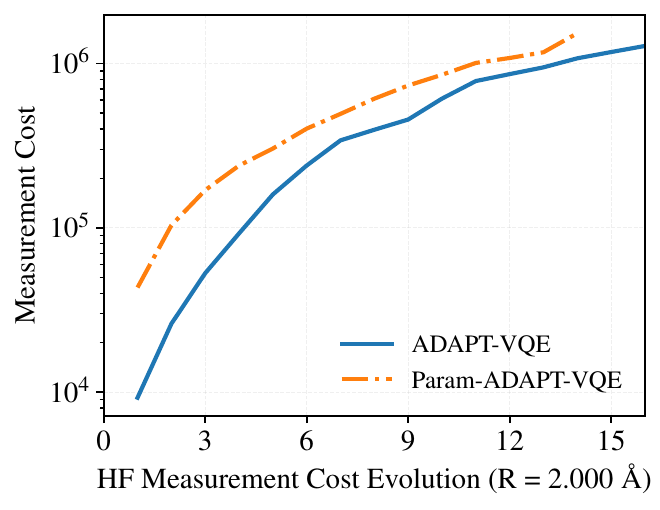}
	\caption{Evolution of energy error and measurement cost with iteration for  HF  at different bond lengths. \label{HF}
	}
\end{figure*}

\begin{figure*}
	\includegraphics[scale=0.3]{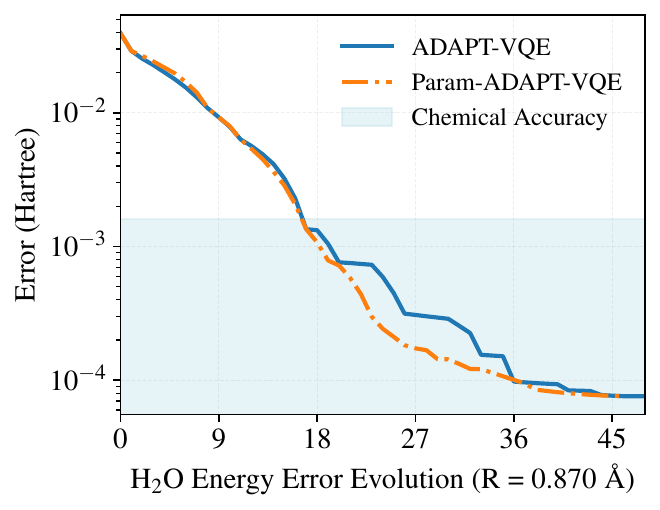}
	\includegraphics[scale=0.3]{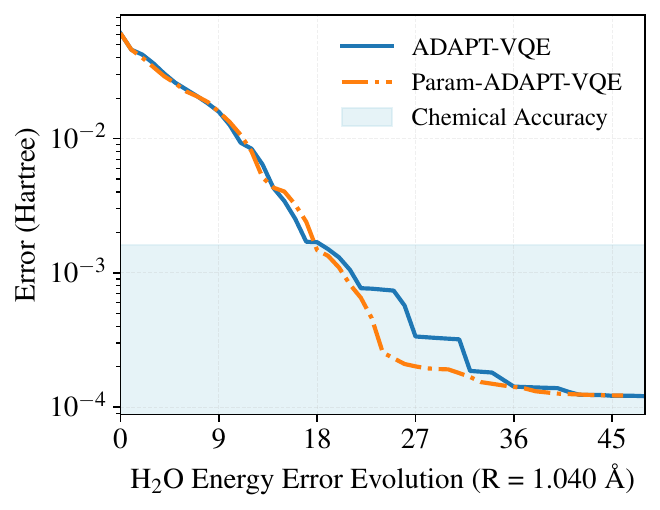}
	\includegraphics[scale=0.3]{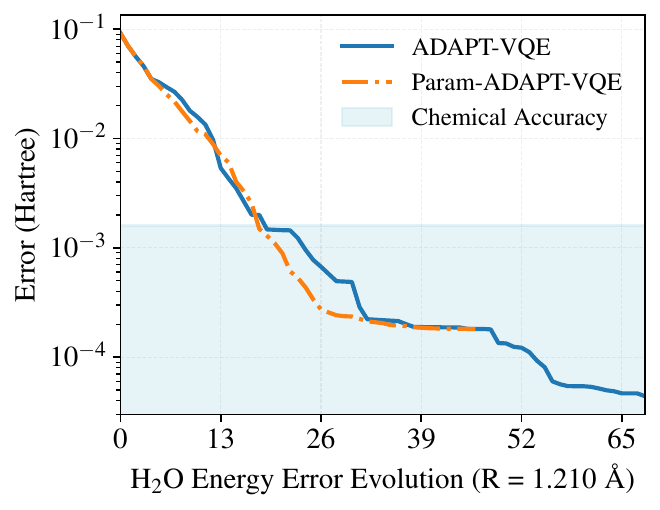}
	\includegraphics[scale=0.3]{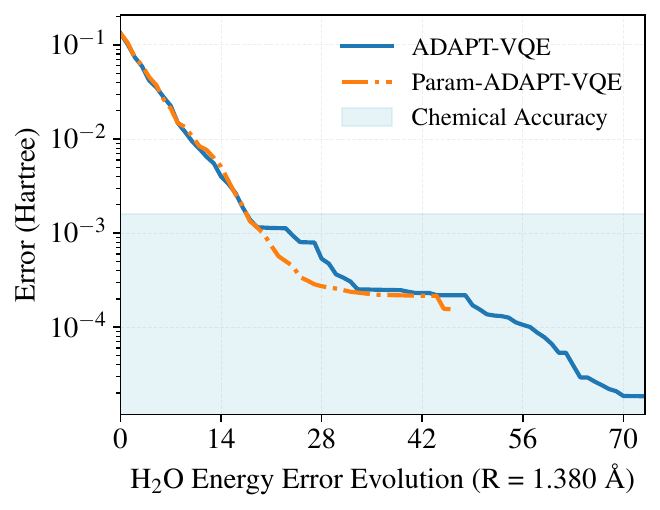}
	\includegraphics[scale=0.3]{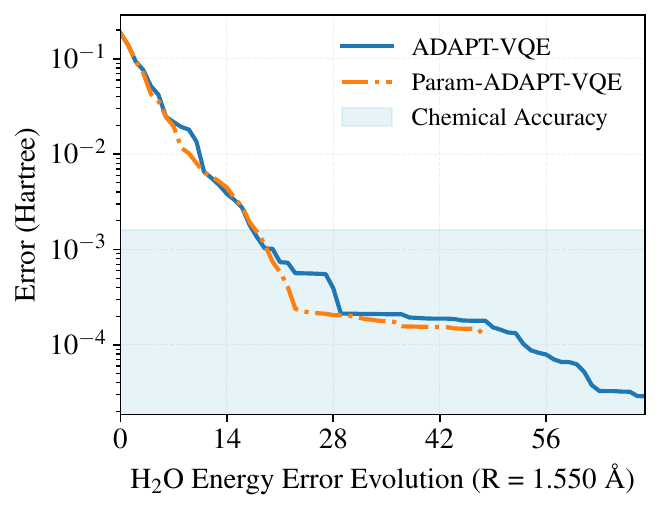}
	\includegraphics[scale=0.3]{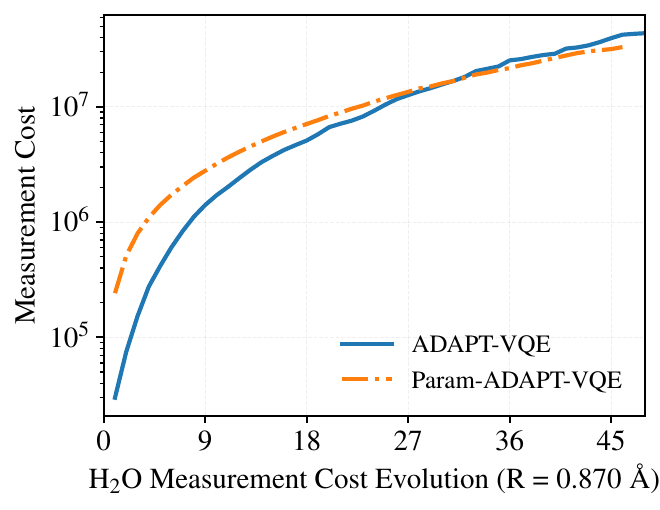}
	\includegraphics[scale=0.3]{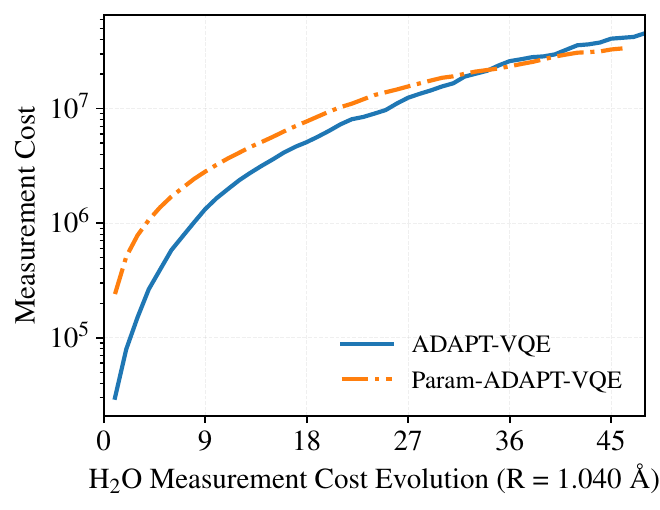}
	\includegraphics[scale=0.3]{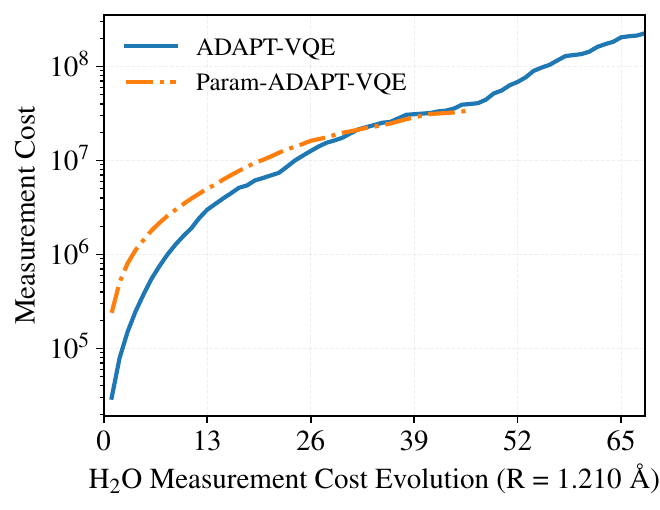}
	\includegraphics[scale=0.3]{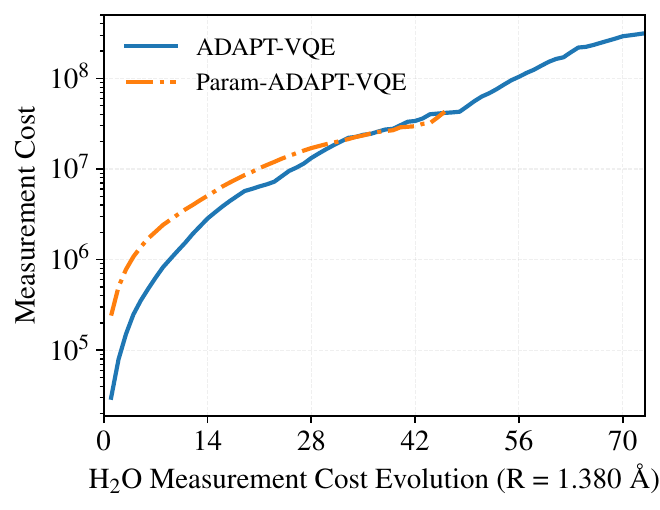}
	\includegraphics[scale=0.3]{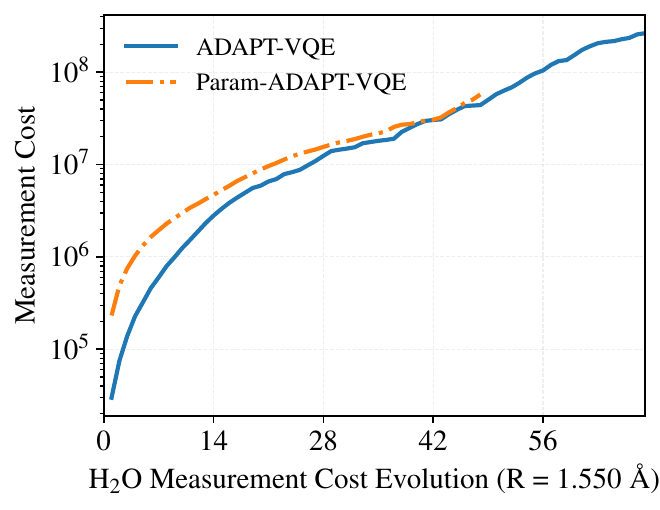}
	\includegraphics[scale=0.3]{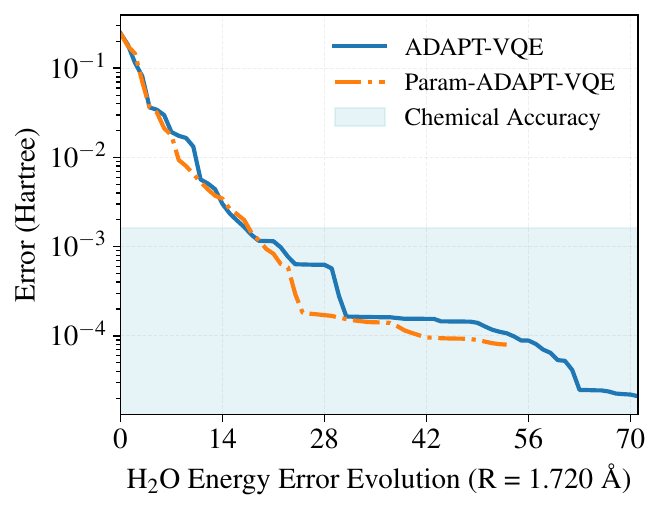}
	\includegraphics[scale=0.3]{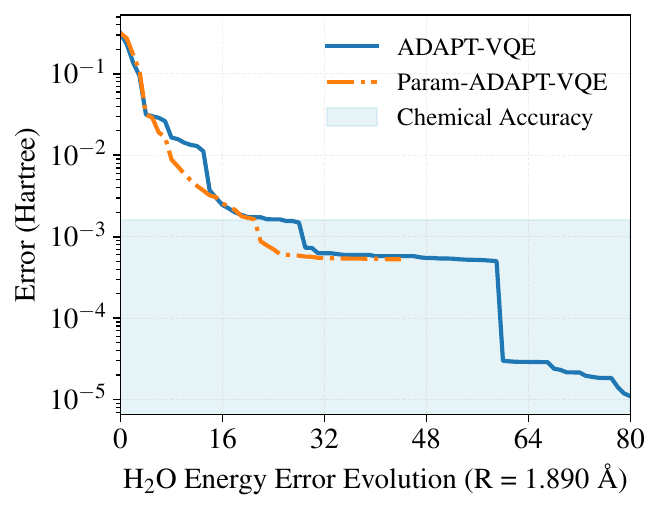}
	\includegraphics[scale=0.3]{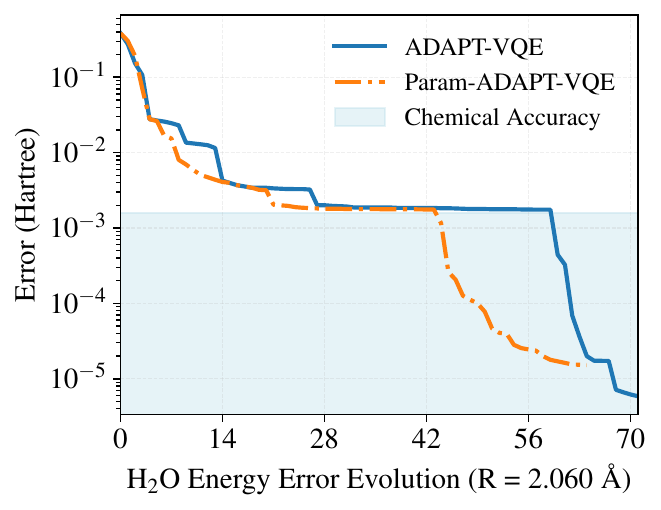}
	\includegraphics[scale=0.3]{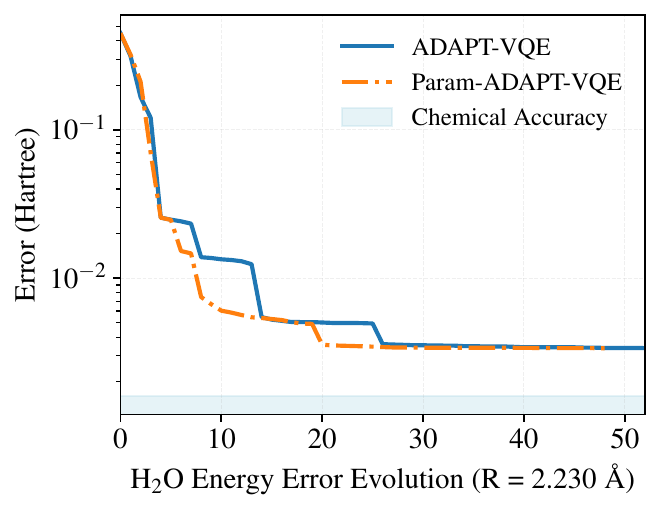}
	\includegraphics[scale=0.3]{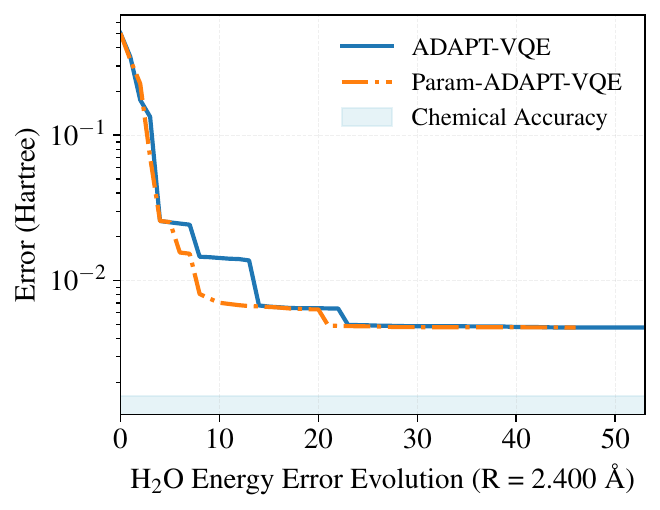}
	\includegraphics[scale=0.3]{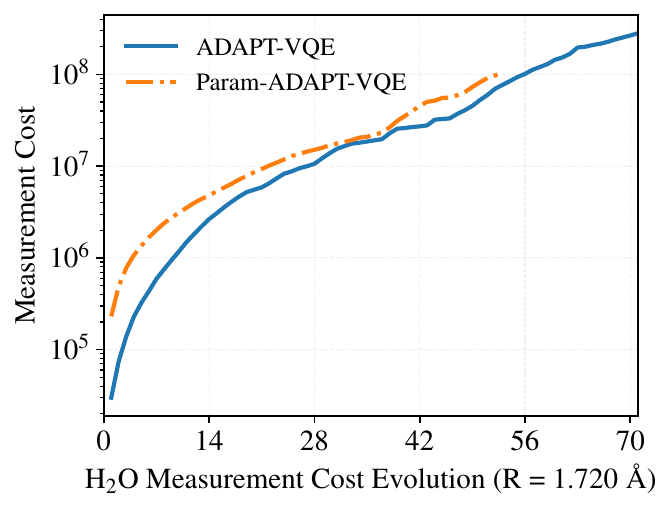}
	\includegraphics[scale=0.3]{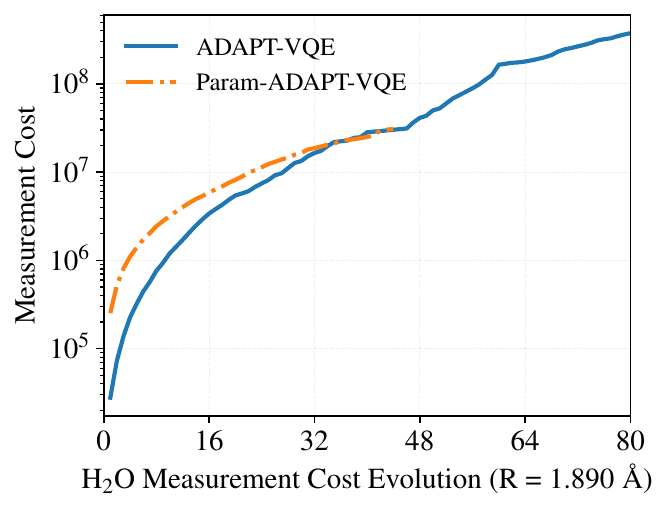}
	\includegraphics[scale=0.3]{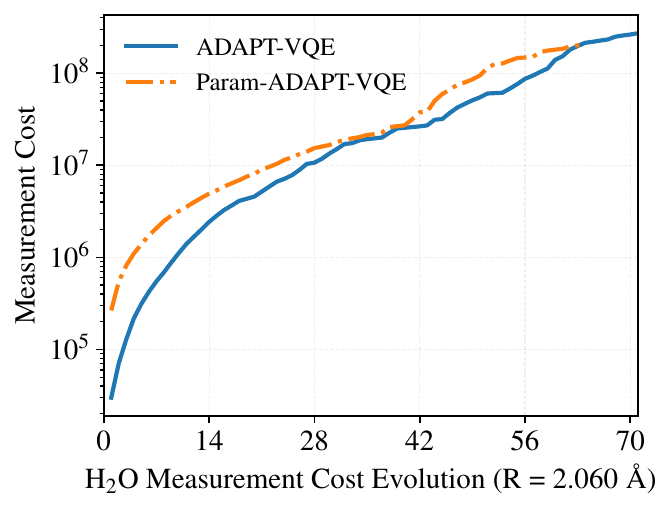}
	\includegraphics[scale=0.3]{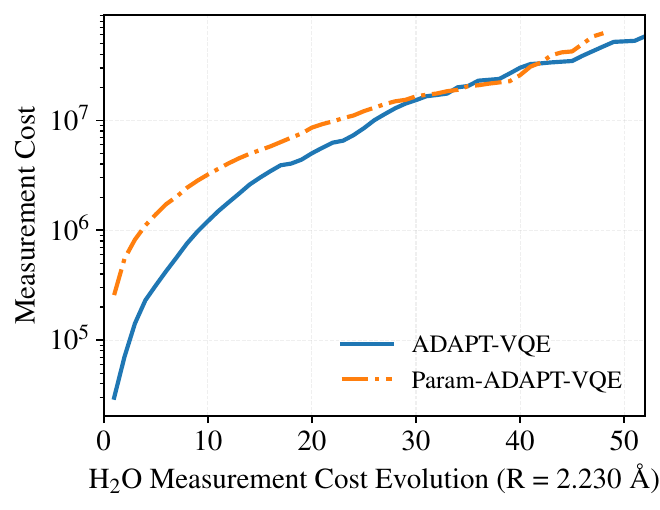}
	\includegraphics[scale=0.3]{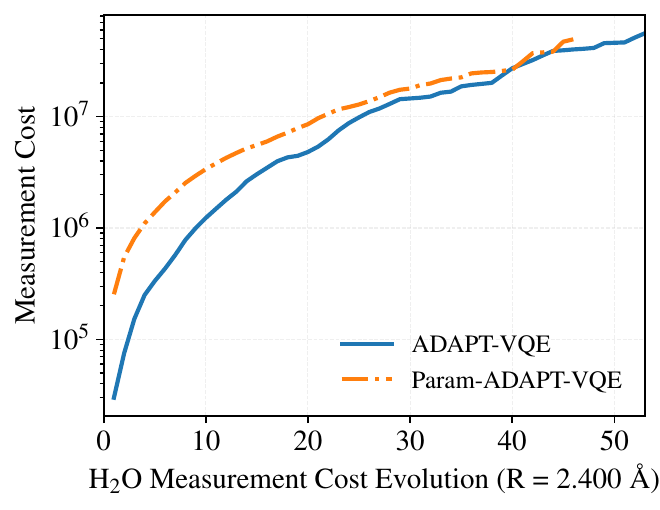}
	\caption{Evolution of energy error and measurement cost with iteration for  \ce{H2O}  at different bond lengths.\label{H2O}
	}
\end{figure*}

\begin{figure*}
	\includegraphics[scale=0.3]{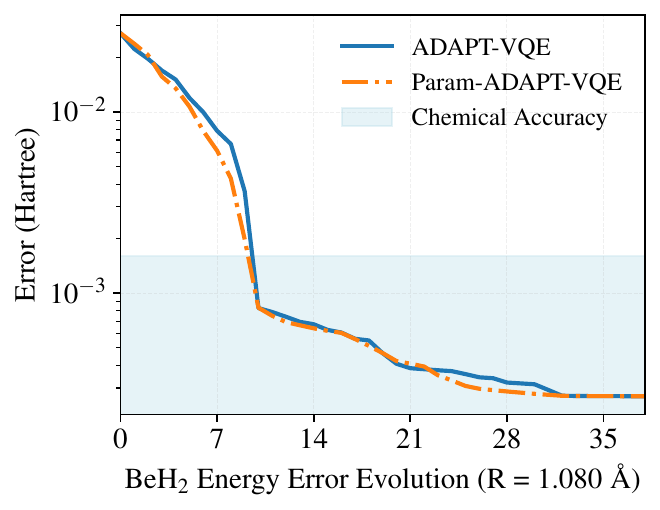}
	\includegraphics[scale=0.3]{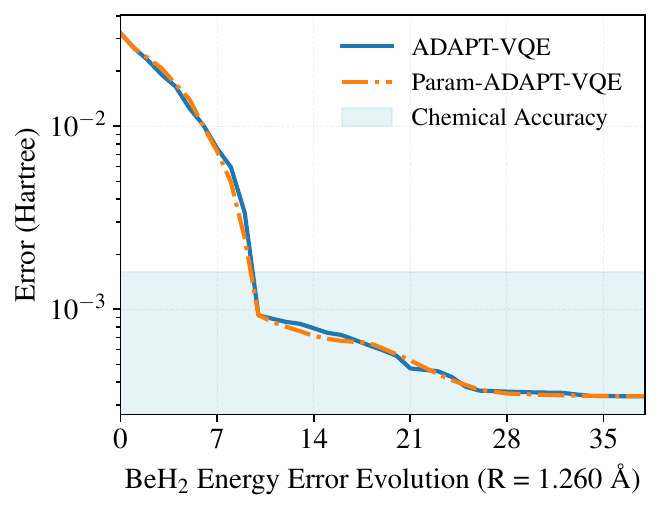}
	\includegraphics[scale=0.3]{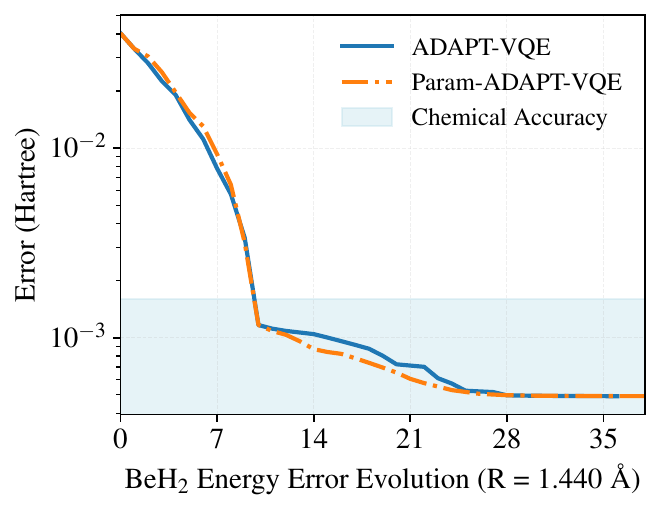}
	\includegraphics[scale=0.3]{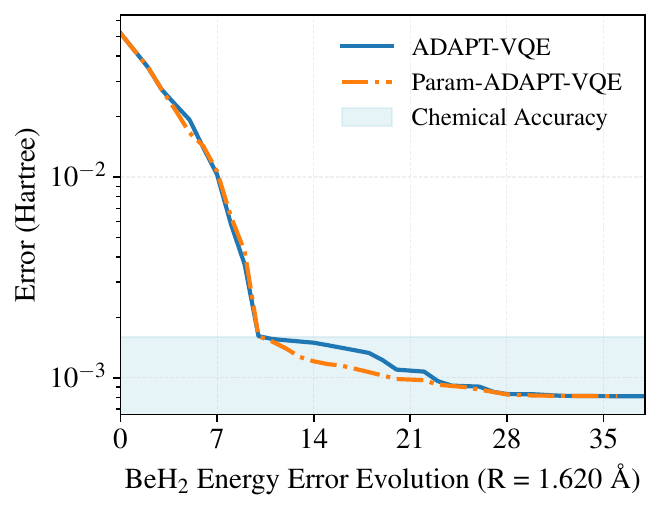}
	\includegraphics[scale=0.3]{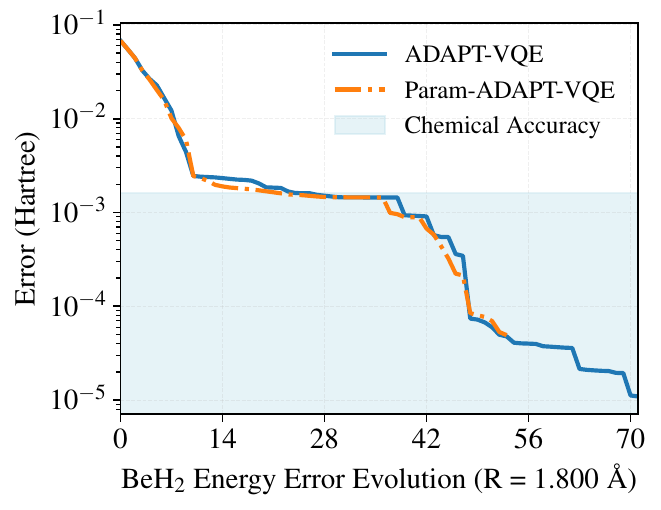}
	\includegraphics[scale=0.3]{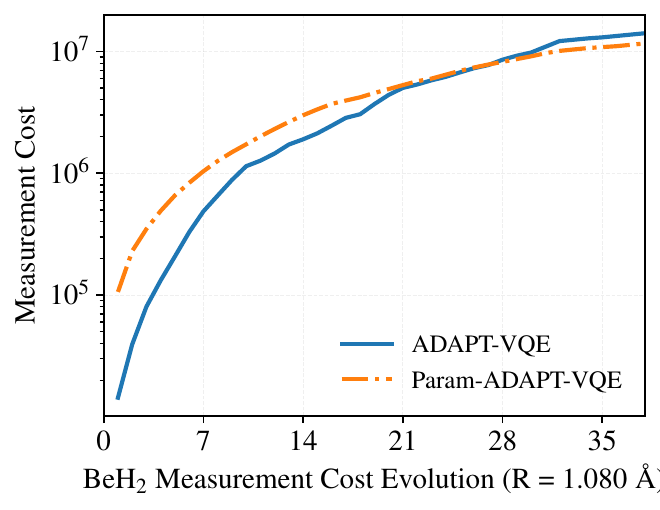}
	\includegraphics[scale=0.3]{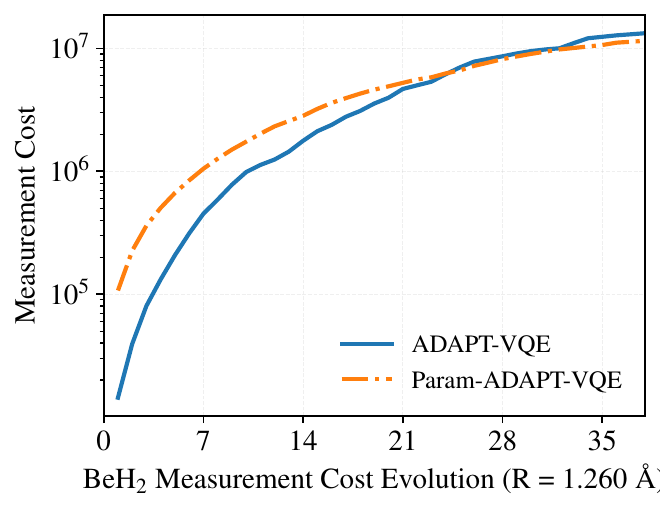}
	\includegraphics[scale=0.3]{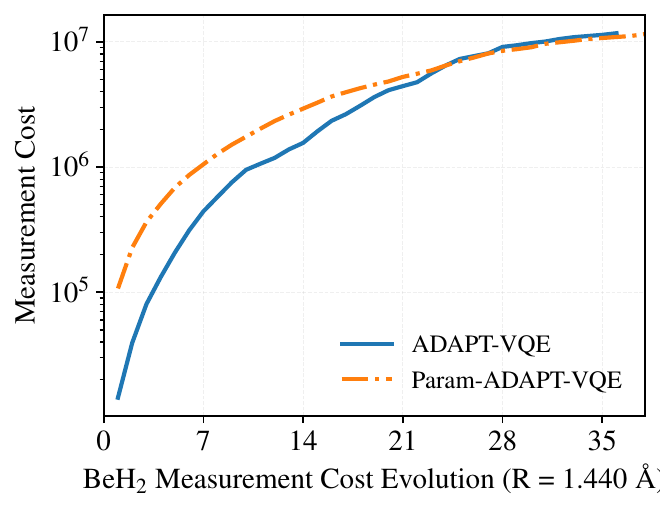}
	\includegraphics[scale=0.3]{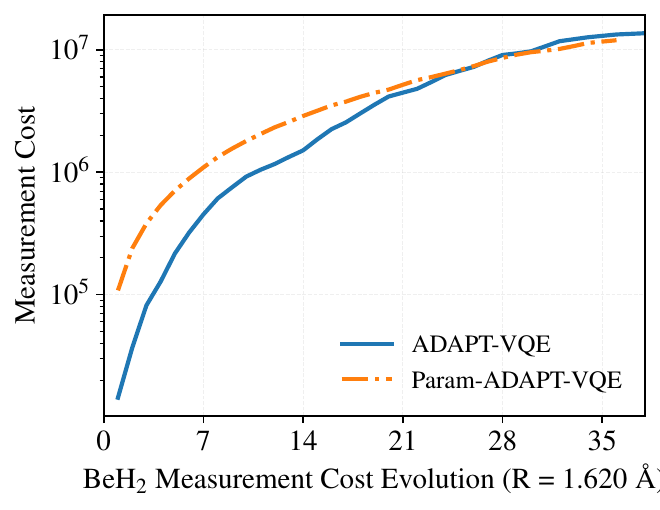}
	\includegraphics[scale=0.3]{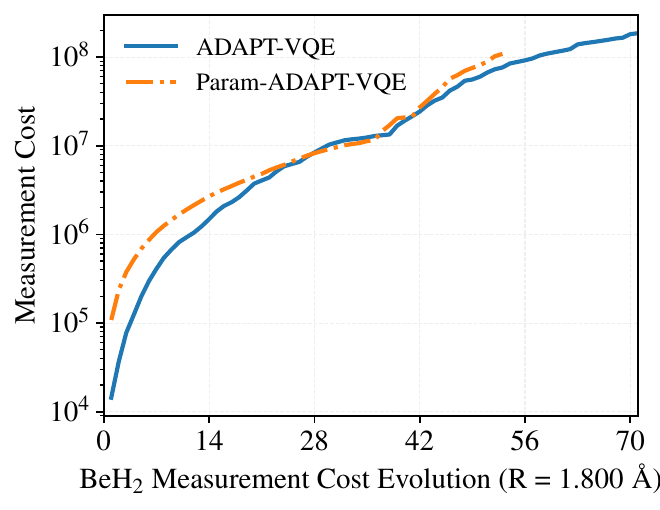}
	\includegraphics[scale=0.3]{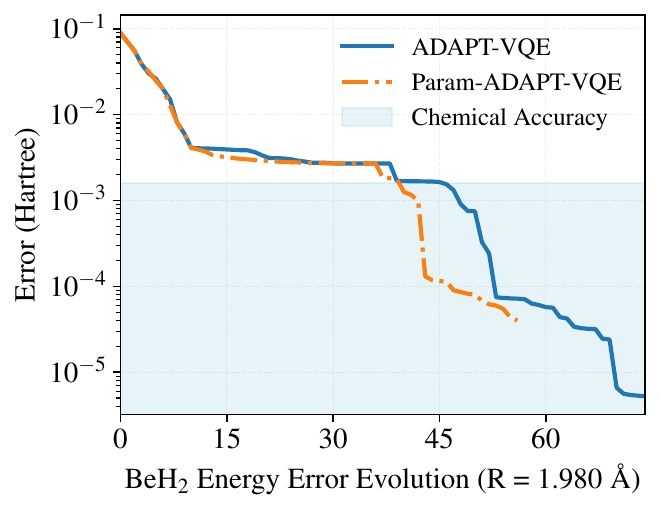}
	\includegraphics[scale=0.3]{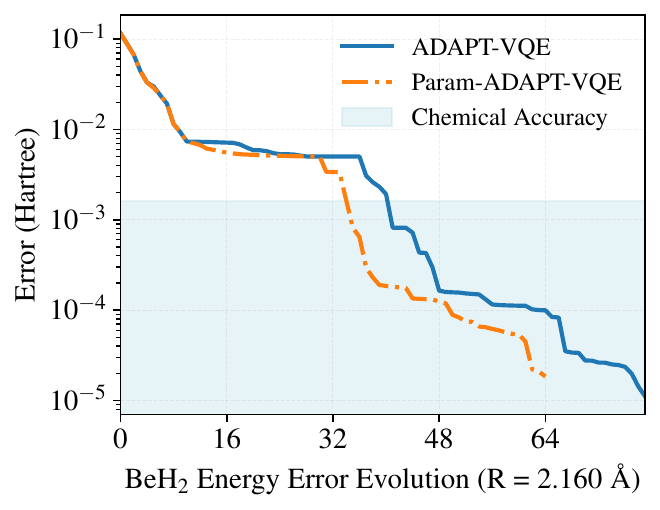}
	\includegraphics[scale=0.3]{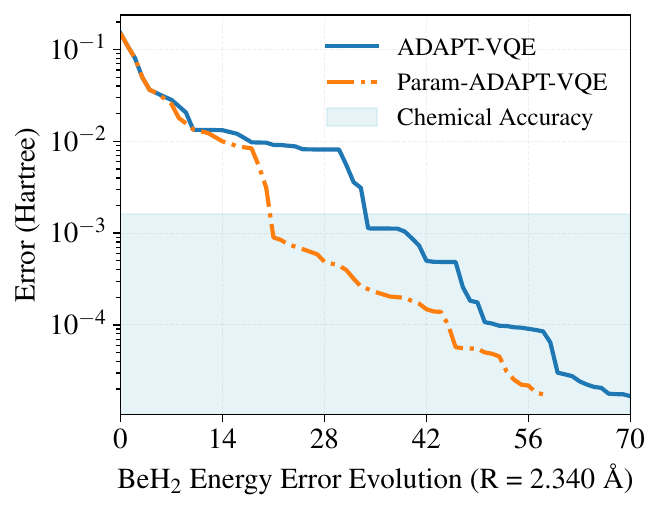}
	\includegraphics[scale=0.3]{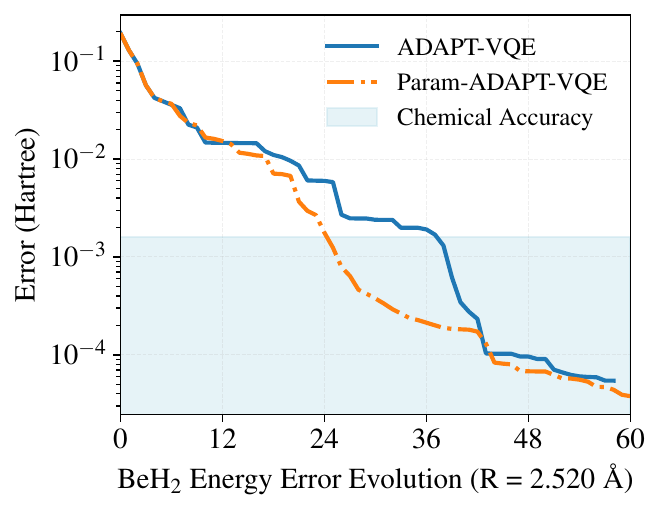}
	\includegraphics[scale=0.3]{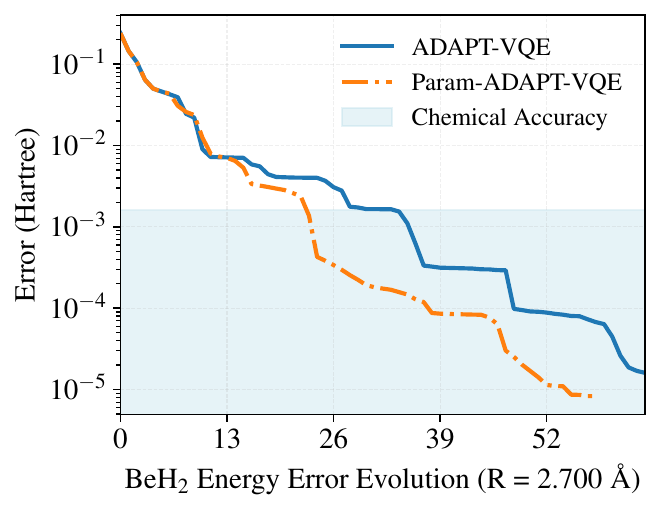}
	\includegraphics[scale=0.3]{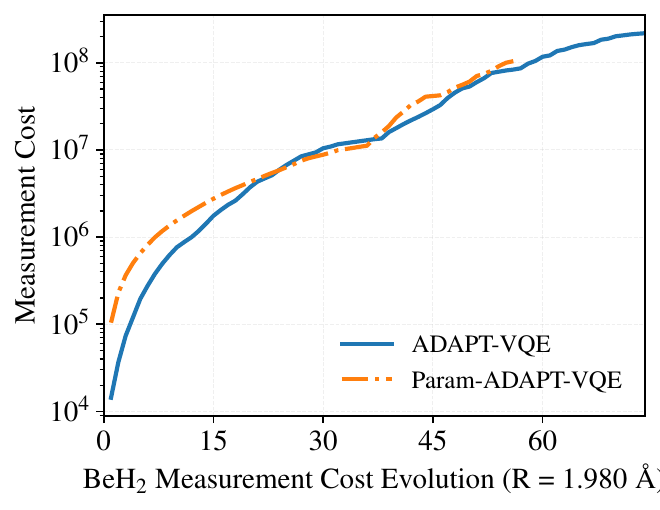}
	\includegraphics[scale=0.3]{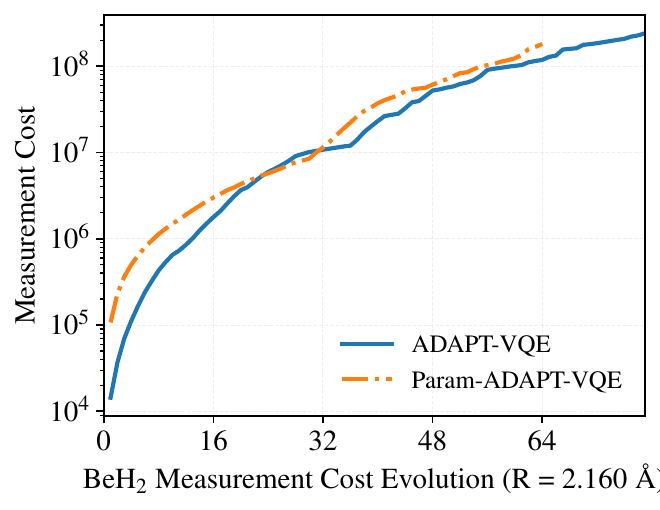}
	\includegraphics[scale=0.3]{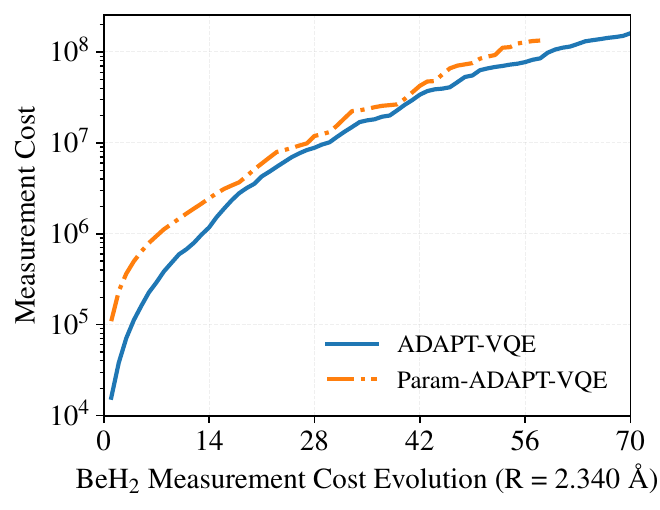}
	\includegraphics[scale=0.3]{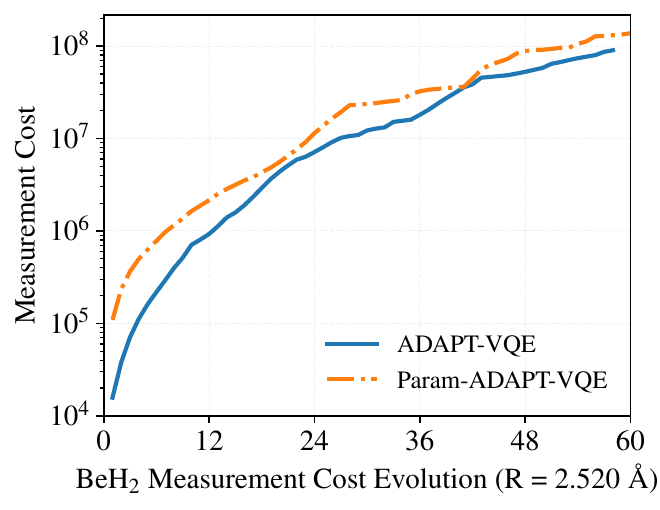}
	\includegraphics[scale=0.3]{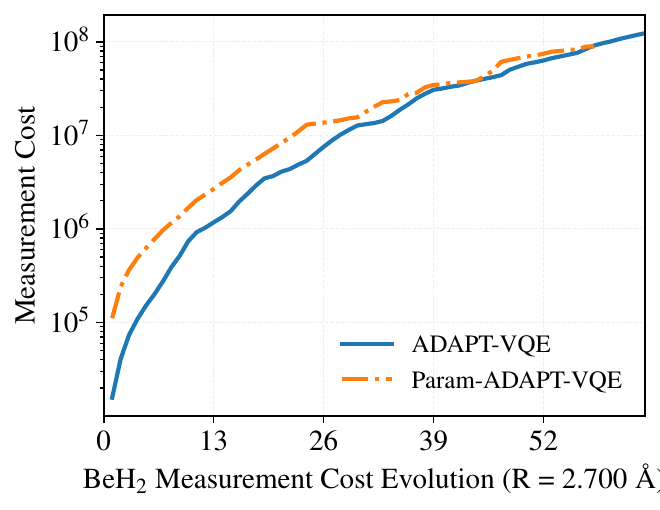}
	\caption{Evolution of energy error and measurement cost with iteration for  \ce{BeH2}  at different bond lengths. \label{BeH2}
	}
\end{figure*}

\begin{figure*}
	\includegraphics[scale=0.3]{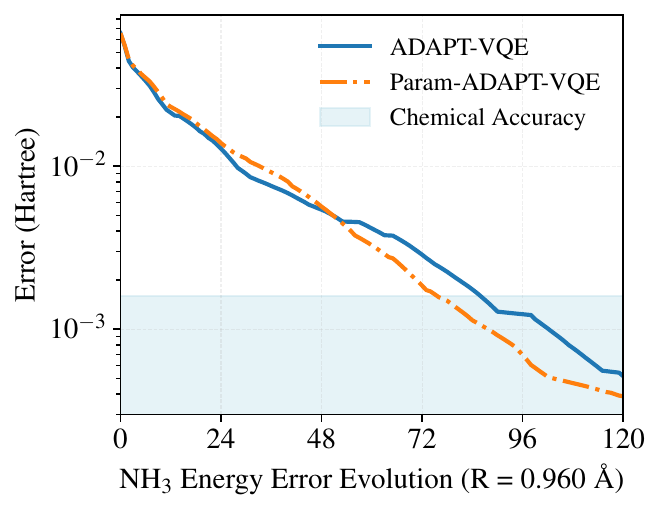}
	\includegraphics[scale=0.3]{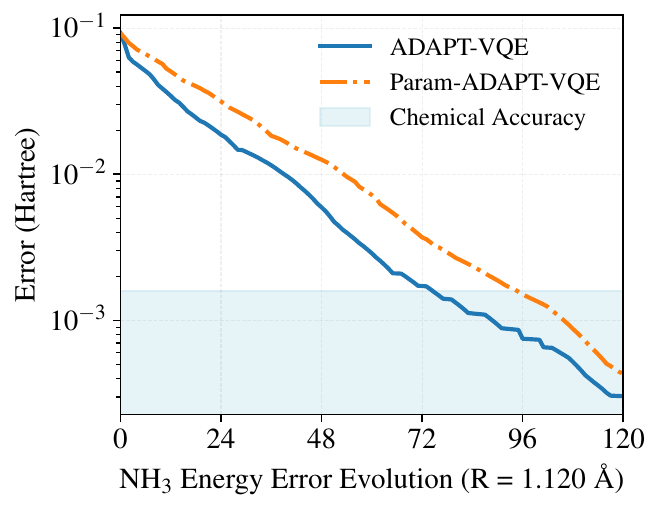}
	\includegraphics[scale=0.3]{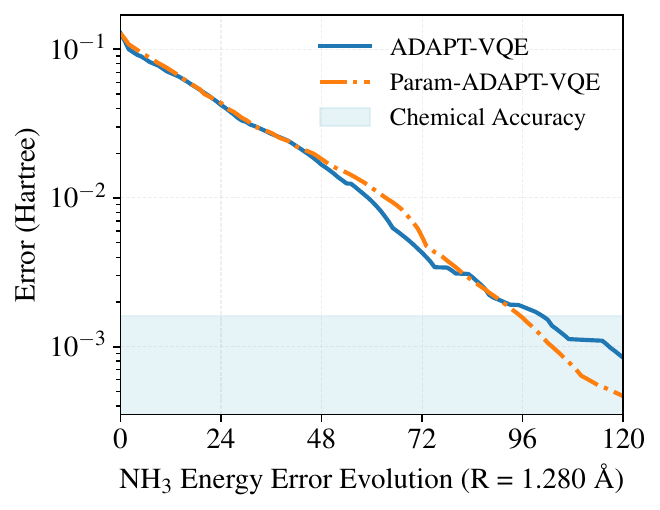}
	\includegraphics[scale=0.3]{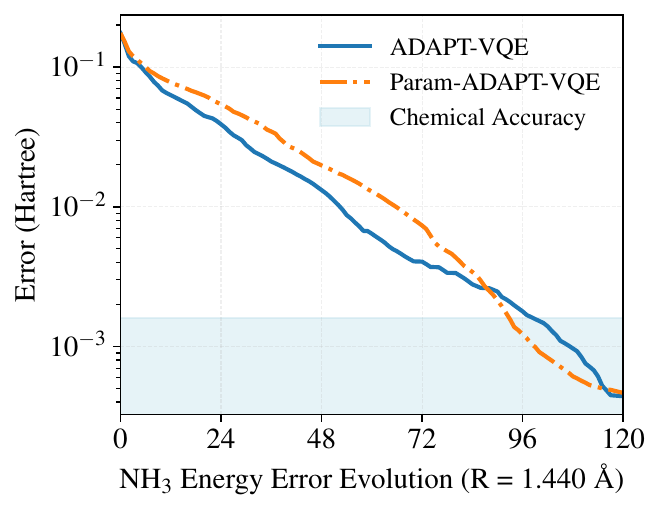}
	\includegraphics[scale=0.3]{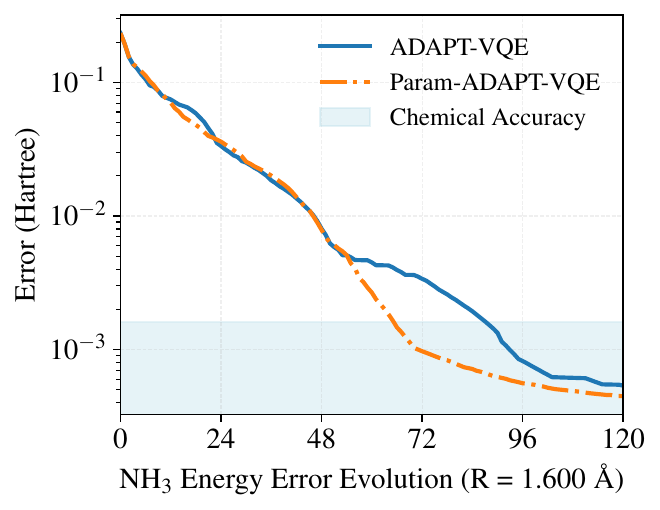}
	\includegraphics[scale=0.3]{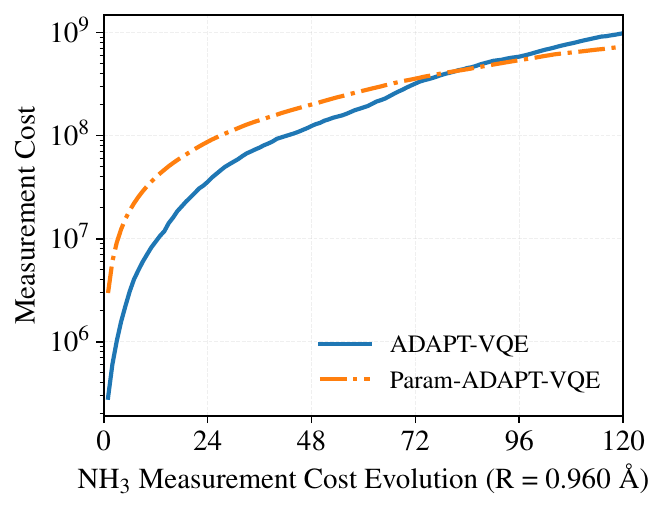}
	\includegraphics[scale=0.3]{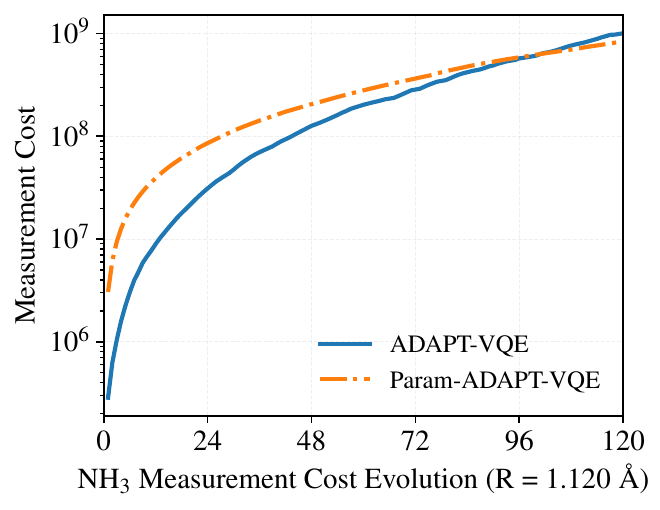}
	\includegraphics[scale=0.3]{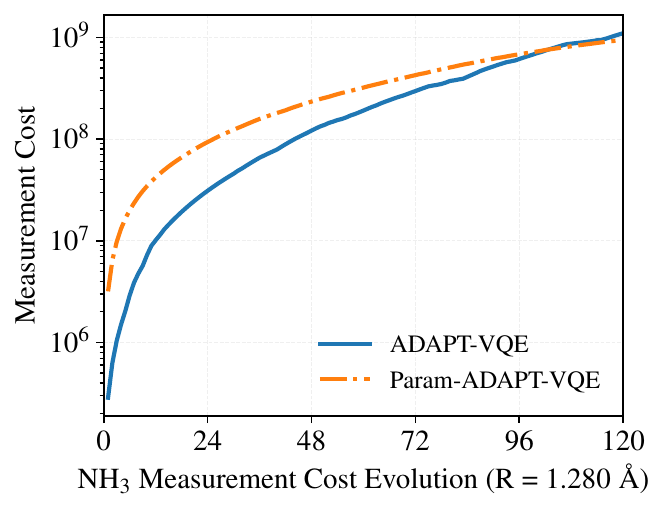}
	\includegraphics[scale=0.3]{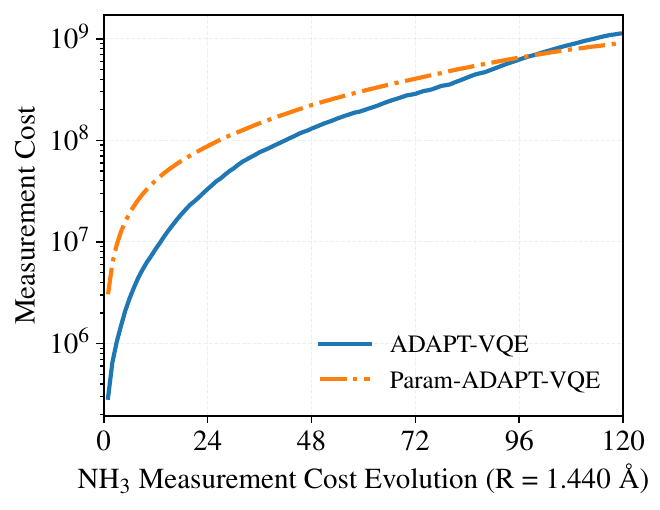}
	\includegraphics[scale=0.3]{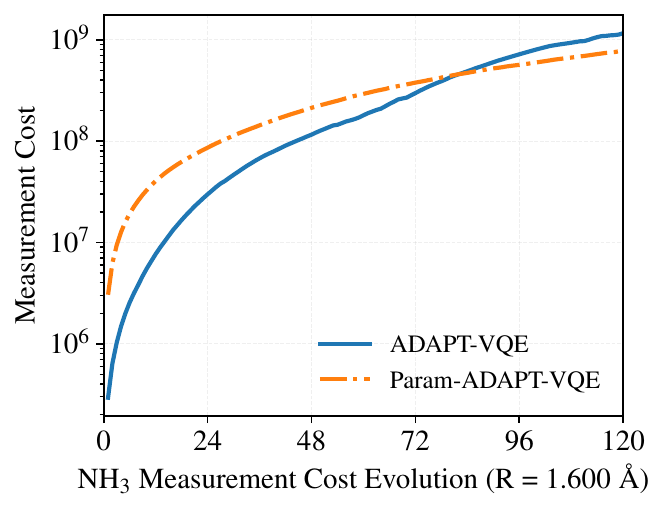}
	\includegraphics[scale=0.3]{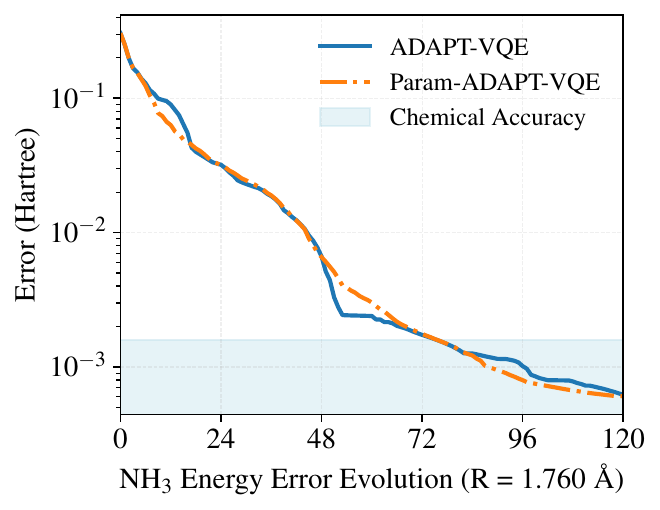}
	\includegraphics[scale=0.3]{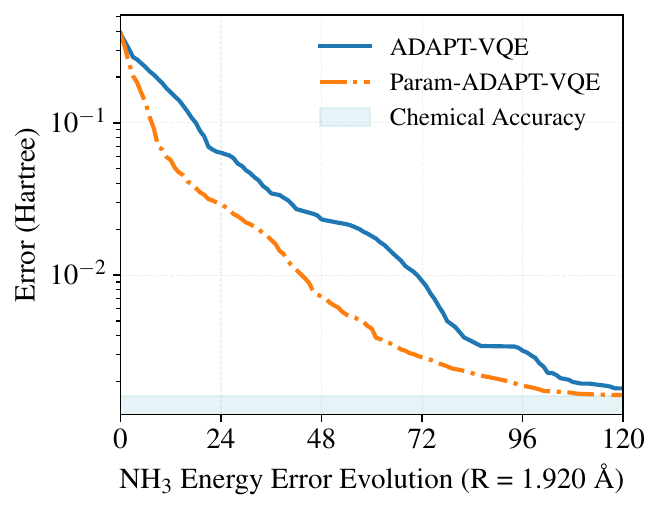}
	\includegraphics[scale=0.3]{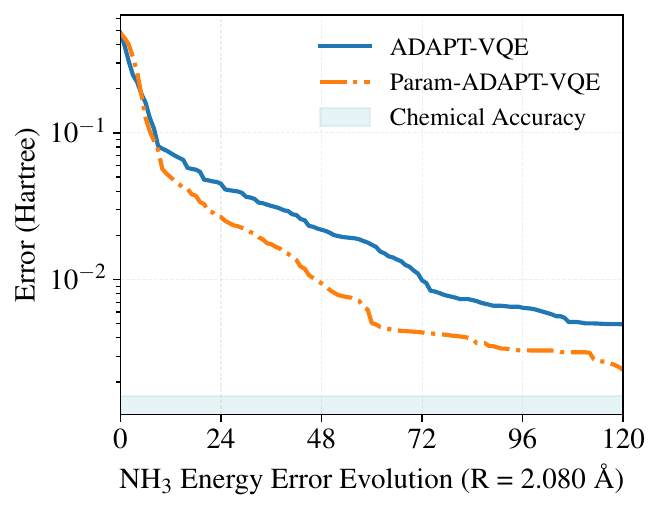}
	\includegraphics[scale=0.3]{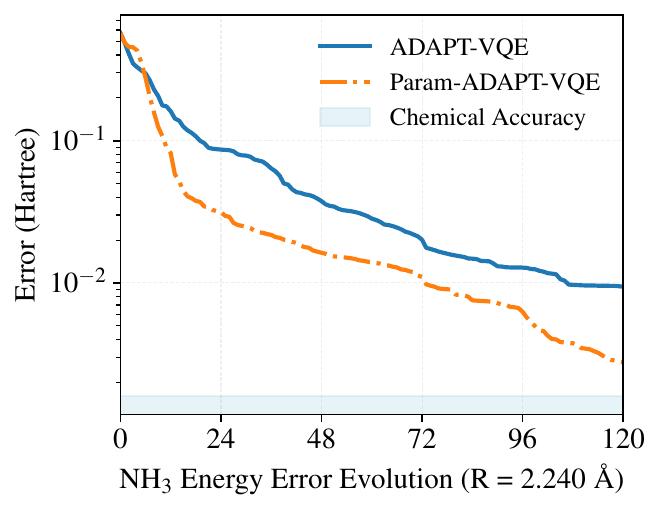}
	\includegraphics[scale=0.3]{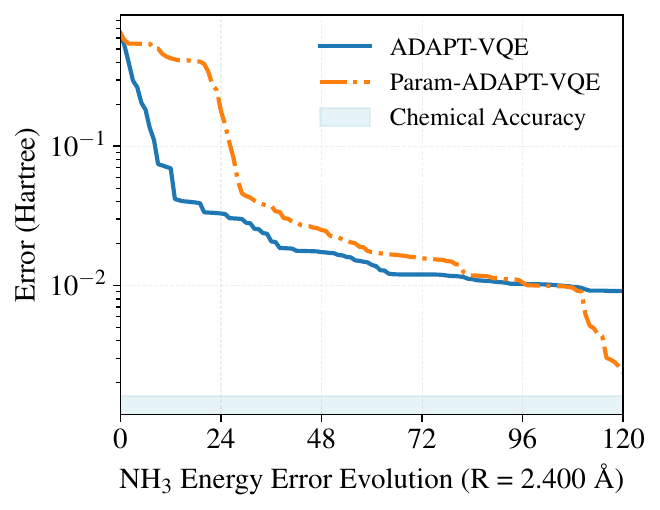}
	\includegraphics[scale=0.3]{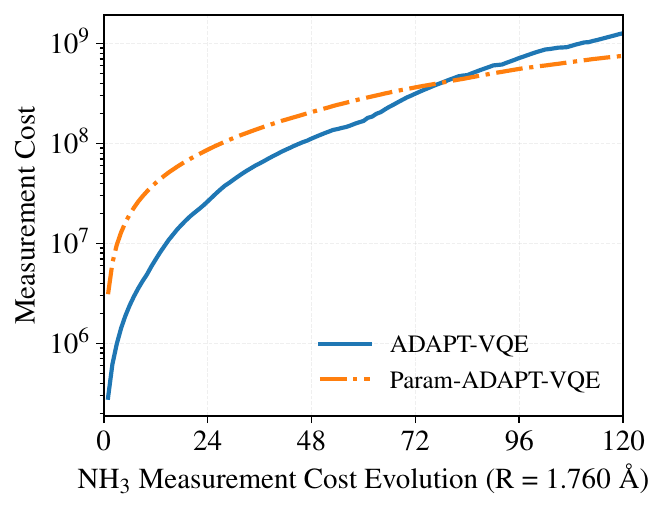}
	\includegraphics[scale=0.3]{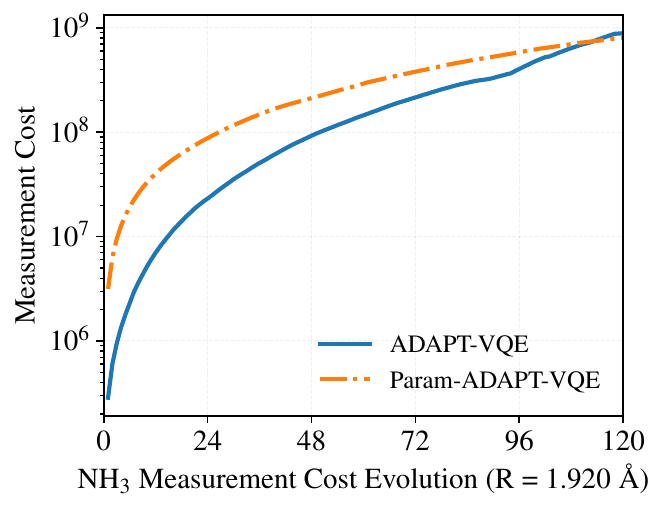}
	\includegraphics[scale=0.3]{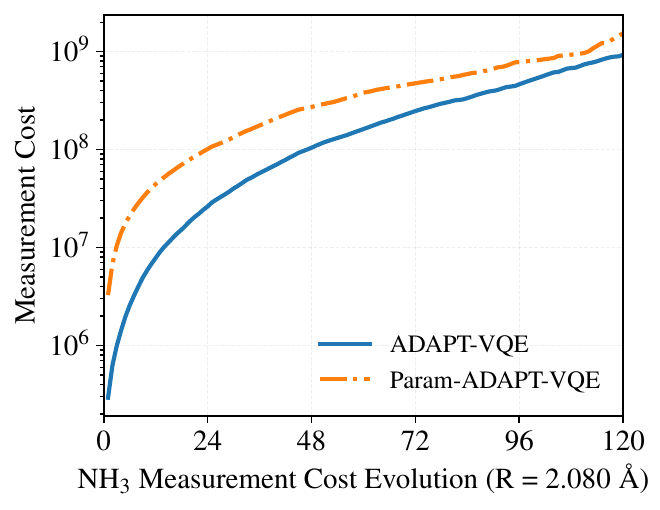}
	\includegraphics[scale=0.3]{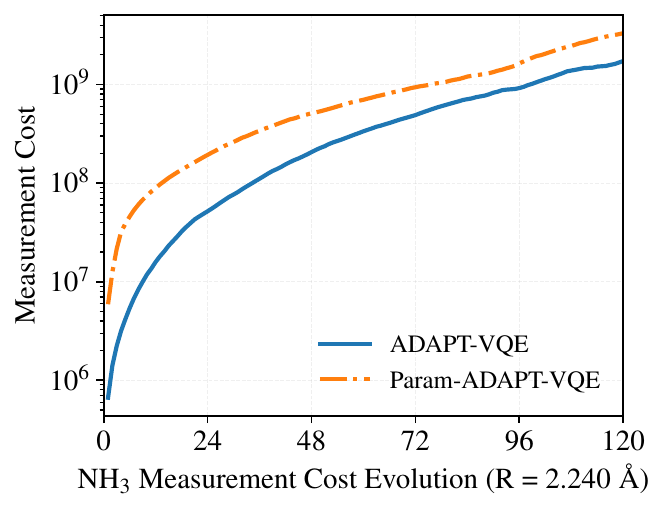}
	\includegraphics[scale=0.3]{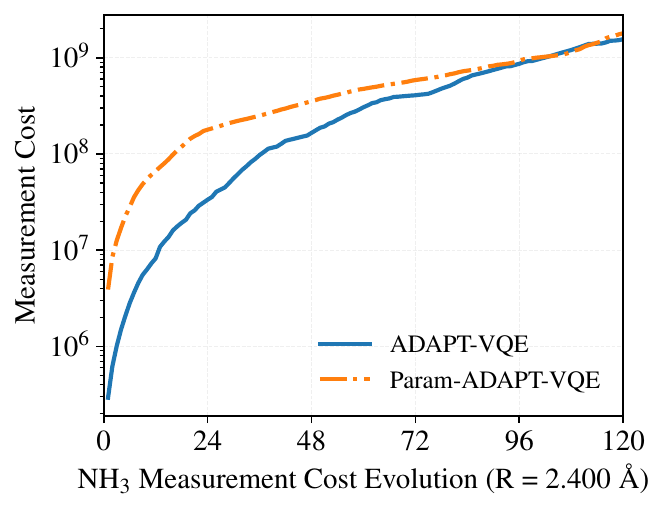}
	\caption{Evolution of energy error and measurement cost with iteration for  \ce{NH3}  at different bond lengths. \label{NH3}
	}
\end{figure*}

To further validate the robustness of our findings, this appendix presents a comprehensive performance comparison between Param-ADAPT-VQE and ADAPT-VQE for \ce{H4} (Figure~\ref{H4}), \ce{LiH} (Figure~\ref{LiH}), \ce{HF} (Figure~\ref{HF}), \ce{H2O} (Figure~\ref{H2O}), \ce{BeH2} (Figure~\ref{BeH2})  and \ce{NH3} (Figure~\ref{NH3}) across a range of bond lengths. The key conclusions drawn from these supplementary analyses are generally consistent with the core results reported in the main text.

\bibliography{References_library}




\end{document}